\documentclass[prd,twocolumn,aps,preprint numbers,superscriptaddress,nofootinbib,showpacs]{revtex4-1}
 \usepackage{amsmath}
\usepackage{amssymb}
\usepackage{graphicx}
\usepackage{color}
\usepackage{bm}
\usepackage{times}
\usepackage{hyperref}
\hypersetup{
  colorlinks=true,
  citecolor=blue,
  linkcolor=blue,
  urlcolor=blue}

\usepackage{epsfig}
\usepackage{dcolumn}
\usepackage{float}

\usepackage{epsfig}
\usepackage{dcolumn}
\def\be{\begin{equation}}
\def\ee{\end{equation}}
\def\bea{\begin{eqnarray}}
\def\eea{\end{eqnarray}}\def\nn{\nonumber}
\def\gsim{\ \rlap{\raise 2pt\hbox{$>$}}{\lower 2pt \hbox{$\sim$}}\ }
\def\lsim{\ \rlap{\raise 2pt\hbox{$<$}}{\lower 2pt \hbox{$\sim$}}\ }
\def\dslash{\kern-4pt \not{\hbox{\kern-2pt $\partial$}}}
\def\pslash{\not{\hbox{\kern-2pt p}}}

\begin{document}

\renewcommand{\arraystretch}{3}
\DeclareGraphicsExtensions{.eps,.ps}


\title{Understanding the Masses and Mixings of One-Zero Textures in 3+1 Scenario 
}


\author{Newton Nath}
\email[Email Address: ]{newton@prl.res.in}
\affiliation{
Physical Research Laboratory, Navrangpura,
Ahmedabad 380 009, India}
\affiliation{Indian Institute of Technology, Gandhinagar, Ahmedabad--382424, India}

\author{Monojit Ghosh}
\email[Email Address: ]{monojit@prl.res.in}
\affiliation{
Physical Research Laboratory, Navrangpura,
Ahmedabad 380 009, India}
        
\author{Shivani Gupta}
\email[Email Address: ]{shivani.gupta@adelaide.edu.au}
\affiliation{
Center of Excellence in Particle Physics (CoEPP), University of Adelaide, Adelaide SA 5005, Australia \label{addr3}            
}

\begin{abstract}

We present a detailed analysis and phenomenological consequences of neutrino mass matrix, $M_{\nu}$, with one-zero texture in the flavor basis where the 
active neutrino sector
is extended by one sterile neutrino (3+1 case). 
In particular, our aim is to explore behaviour of the sterile mixing parameters in detail when one of the elements of the neutrino mass matrix goes to zero.
To study this, we consider two distinct mass spectrum of the active neutrinos: (i) completely hierarchical mass spectrum with a vanishing neutrino mass
and (ii) completely quasidegenerate mass spectrum.
In 3+1 scenario, the low energy neutrino mass matrix, $M_{\nu}$, is a $4 \times 4$ matrix and has 10 independent elements.
Thus it can have 10 possible one-zero textures. From the earlier studies it can be inferred that,
 if one assumes one vanishing neutrino mass, then 
only seven of these $M_{\nu}$ are phenomenologically allowed by the current
neutrino oscillation data. On the other hand, if the neutrinos are quasidegenerate then there are eight phenomenologically viable one-zero textures.
In this present work, we study the correlations between the sterile mixing parameters for each of these allowed textures for both mass spectrum
and also their implications on the effective Majorana mass.  

\end{abstract}
\preprint{ADP-15 - 47 / T949}
\maketitle

\section{Introduction}

Compelling evidence have been provided by solar, atmospheric, reactor and accelerator neutrinos
for the existence of neutrino oscillations among the three active neutrinos. 
Neutrino oscillation in three generation can be described by 
three mixing angles: $\theta_{12}$, $\theta_{23}$ and $\theta_{13}$, two mass squared differences:
$\Delta m_{21}^2(m_2^2 - m_1^2)$ and $\Delta m_{31}^2(m_3^2 - m_1^2)$ and one Dirac type CP phase $\delta_{CP}$. 
There are two additional phases if neutrinos are Majorana particles.
The global fits of neutrino oscillation data in the
framework of 3 $\nu$ mixing \cite{Gonzalez-Garcia:2015qrr,Forero:2014bxa,Capozzi:2013csa} give us precise values of the neutrino oscillation parameters.
At present, the major unknowns in the three flavour oscillation picture are: (i) the
sign of the atmospheric mass squared difference $\Delta m^2_{31}$ i.e., whether
the neutrino mass hierarchy is normal or inverted ($\Delta m^2_{31} > 0$: NH or $\Delta m^2_{31} < 0$ : IH), 
(ii) the octant of $\theta_{23}$ i.e., lower or higher ($\theta_{23} < 45^\circ$: LO or $\theta_{23} > 45^\circ$: HO) and (iii) the precise value
of the leptonic CP phase $\delta_{CP}$. 
There are various current ongoing / future upcoming neutrino oscillation experiments
 dedicated to the  
determination of the remaining unknown oscillation parameters.


Another intriguing aspect of current neutrino physics is the existence of light sterile neutrino.
Sterile neutrinos are $SU(2)$ singlet which implies that they do not 
take part in the usual weak interactions. 
However, they can
mix with the active neutrinos and thus probed in the neutrino oscillation
experiments. The oscillation results of the LSND experiment
shows the evidence that there should be at least one sterile neutrino having mass 
in the $\sim$ eV scale \cite{Athanassopoulos:1996jb,Athanassopoulos:1997pv,Aguilar:2001ty}. The latest
antineutrino data of MiniBooNE also have some overlap with the
the allowed regions of the LSND experiment supporting the sterile neutrino hypothesis \cite{Aguilar-Arevalo:2013pmq}.
\footnote{Note that the excess of the MiniBooNE neutrino data marginally agrees
with a simple two neutrino oscillation formalism and requires more data to resolve the issue.}
In addition, studies 
of reactor antineutrino spectra show a 3\% enhancement in the fluxes as compared to the previous calculation. 
With these new re-evaluated fluxes, the ratio of observed event rate to predicted rate for $< 100$ m reactor experiments shifts from 0.976 to 0.943, 
giving rise to reactor
neutrino anomaly \cite{Mention:2011rk}. This deficit can not be explained in three flavour 
framework and evokes the presence of sterile neutrinos.
The recently observed Gallium anomaly also seems to support the sterile neutrino 
hypothesis \cite{Giunti:2010zu}. 
 Although it is possible to fit experimental data with 
more than one light sterile neutrinos, the 3+1 hypothesis i.e., three active neutrinos 
in the sub-eV scale and one eV scale sterile neutrino, is
considered to be minimal and still not completely ruled out from the cosmological observations. 
Recent analysis of the Planck data shows that the possibility of light sterile neutrino is ruled at $3 \sigma$ within the $\Lambda$CDM model. However it is possible to have one light sterile neutrino in the eV scale if one deviates slightly from the base $\Lambda$CDM model \cite{Ade:2015xua}.   
The possibility of having more than one light sterile neutrino 
is highly disfavoured in cosmology. 
In the 3+1 model, there are total six mixing angles, three Dirac phases, three Majorana phases and three mass squared differences. The new mixing angles, which arise due to the inclusion of one sterile neutrino,
are $\theta_{14}$, $\theta_{24}$ and $\theta_{34}$. The additional mass squared
difference is $\Delta m^2_{LSND}$ which can be either $\Delta m^2_{41}$ if the active neutrinos are normal hierarchical or $\Delta m^2_{43}$ when the active neutrinos are inverted hierarchical. 

Neutrino mass matrices provide important tool for the investigation of underlying
symmetries and resulting dynamics. The first step in this direction is to construct the
neutrino mass matrix in the flavor basis. However, the reconstruction results in a large
variety of possible structures of the mass matrices depending strongly on the mass scale, mass
hierarchy and the Majorana phases. Several proposals have been made in literature
to restrict the form of neutrino mass matrix and to reduce the number of free parameters
one of which is the zero texture. Zero texture implies that one or more elements are
relatively small compared to others. 
Extensive studies of neutrino mass matrices with zero textures in the standard three neutrino scenario have been done in 
\cite{Frampton:2002yf,Dev:2006qe,Xing:2002ta,Xing:2002ap,Desai:2002sz,Dev:2007fs,Dev:2006xu,Kumar:2011vf,Fritzsch:2011qv,Meloni:2012sx,Ludl:2011vv,Grimus:2012zm}.
These neutrino mass matrix textures, in general, may be obtained by imposing certain Abelian family symmetries \cite{Grimus:2004hf,Low:2005yc,Dev:2011jc}.
For the three generations the maximum number of zero textures in neutrino mass matrix is
two whereas the addition of a sterile neutrino to the standard three neutrino picture (3+1)
can increase the allowed zero textures to three \cite{Ghosh:2012pw,Ghosh:2013nya,Zhang:2013mb}.

In this paper we investigate the phenomenology of the one-zero textures in the low energy neutrino mass matrix for 3+1 scheme. 
In particular, our main focus in this present work is to understand the mixing pattern of the sterile parameters which give rise to allowed one-zero textures.
The study of one-zero texture in 3+1 scheme 
has been done in \cite{Ghosh:2013nya}. The paper mainly studied the nature of the matrix elements as a function of the lowest mass. 
This facilitates to identify the mass ranges over which it is possible to have one-zero texture for a given matrix element. However, in this paper the main emphasis is to
study the correlations between the different active sterile mixing parameters.
To understand this, the strategy of the present work is as follows.
We consider two distinct mass spectrum: (i) the completely hierarchical mass spectrum in which the lowest 
active neutrinos mass is zero \footnote{ The one-zero singular models are found to be quite rich in phenomenology in comparison to non singular models for 3 active neutrinos \cite{Lashin:2011dn}.} 
and (ii) the quasidegenerate mass spectrum where all the three active neutrinos have mass of the same order. 
From the results of \cite{Ghosh:2013nya}, one can infer that for completely hierarchical mass spectrum the number of allowed texture is seven and if one considers quasidegenerate spectrum then eight one-zero textures become allowed. In this paper we study the correlation 
between different sterile mixing parameters for every allowed one-zero textures corresponding to the above two mass spectrum. These results are important to understand the underlying parameter space of the neutrino mass matrix and they can also help in putting constraint on the free parameters of the
models involving light sterile neutrinos \cite{Barry:2011wb,Zhang:2011vh,Dev:2012bd}. 

The plan of the paper goes as follows. In Section \ref{sec2}, we present the mass and mixing patterns along with the experimental constraints on the neutrino oscillation parameters when a sterile neutrino of mass $\sim$ 1 eV is added to standard three neutrino picture. In Section \ref{sec3} we present our formalism for 
obtaining one-zero textures assuming: (a) the lowest neutrino mass to vanish and (b) the masses
of active neutrinos are quasidegenerate i.e., approximately of same order.
In Section \ref{sec4} we study the phenomenological consequences of all the ten elements of the mass matrix
for the above mentioned scenarios.
We also present the predictions on effective Majorana mass, governing neutrinoless double beta decay
for all the phenomenologically viable textures. 
We finally conclude in Section \ref{sec5}.  

\section{Masses and mixings in 3+1 scenario and experimental constraints}
\label{sec2}

The sterile state of mass $\sim \rm{eV}$ 
can be added to the standard three neutrino mass states
in two different ways. In one case active neutrinos have normal hierarchical mass spectrum i.e.,

SNH: $m_1 \approx m_2 < m_3 < m_4$ which implies, 
\bea \rm
m_2&=&\sqrt{m_1^2+\Delta m_{21}^2}, \\
m_3&=& \sqrt{m_1^2+\Delta m_{21}^2+\Delta m_{31}^2}, \\
m_4&=&\sqrt{m_1^2+\Delta m_{41}^2}.
\eea
Here $m_1$, $m_2$ and $m_3$ are active neutrino masses and $m_4$ is the sterile neutrino mass. 

\begin{figure}[h!]
 \begin{center}
 \includegraphics[scale=0.42]{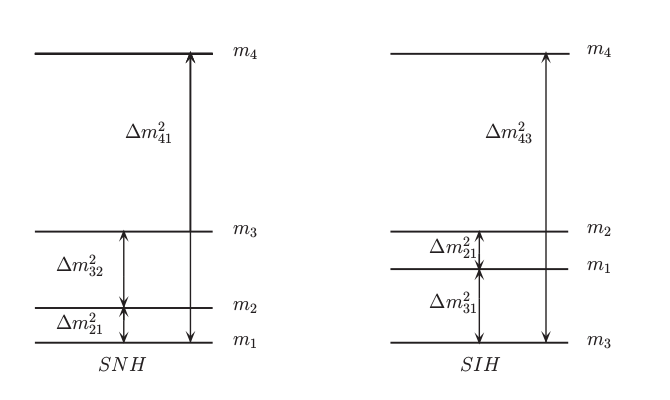}
\end{center}
\vspace{-0.5cm}
\caption{Allowed  mass spectrum in 3+1 scheme for normal (SNH) and inverted (SIH) hierarchy.}
\label{fig1}
 \end{figure}

In the second case the active neutrinos will be inverted hierarchical i.e., \\
 SIH : $m_3 < m_1 \approx m_2 < m_4$ implying
\bea \rm
m_1&=& \sqrt{m_3^2+\Delta m_{31}^2}, \\
m_2&=&\sqrt{m_3^2+\Delta m_{31}^2+\Delta m_{21}^2}, \\
m_4&=&\sqrt{m_3^2+\Delta m_{43}^2}.
\eea
The mass spectrum for SNH and SIH are displayed in Fig. \ref{fig1}. 

\def\arraystretch{1.3}
\begin{table} 
\centering
\begin{tabular}{|c|ccc|}
\hline
Parameter &  Best Fit $\pm 1\sigma$ & $2\sigma$ range & $3\sigma$ range \\
\hline
$\Delta m^2_{21}[10^{-5}~\mathrm{eV}^2] $  & 7.60$_{-0.18}^{+0.19}$ &7.26 -- 7.99 &7.11 -- 8.18 \\
$\Delta m^2_{31}[10^{-3}~\mathrm{eV}^2] $ (NH) & 2.48$_{-0.07}^{+0.05}$ & 2.35 -- 2.59 & 2.30 -- 2.65 \\
$\Delta m^2_{31}[10^{-3}~\mathrm{eV}^2] $ (IH) & 2.38$_{-0.06}^{+0.05}$  & 2.26 -- 2.48 & 2.20 -- 2.54 \\
$\sin^2 \theta_{12}/10^{-1}$  & 3.23$\pm$0.16  & 2.92 -- 3.57 & 2.78 -- 3.75 \\
$\sin^2 \theta_{13}/10^{-2}$ (NH) & 2.26$\pm$0.12 & 1.95 -- 2.74 & 1.90 -- 2.62 \\
$\sin^2 \theta_{13}/10^{-2}$ (IH) & 2.29$\pm$0.12 & 2.02 -- 2.78 & 1.93 -- 2.65 \\
$\sin^2 \theta_{23}/10^{-1}$ (NH)  & 5.67$_{-1.24}^{+0.32}$  & 4.13 -- 6.23  & 3.93 -- 6.43 \\
$\sin^2 \theta_{23}/10^{-1}$ (IH)  & 5.73$_{-0.39}^{+0.25}$  & 4.32 -- 6.21 & 4.03 -- 6.40 \\
$\rm \Delta m_{LSND}^2~(\mathrm{eV}^2)$  &  0.89$_{-0.11}^{+0.16}$ & 0.73 -- 2.05 &0.7 -- 2.5 \\
$ \sin^2\theta_{14} $ & 0.025$_{-0.015}^{+0.025}$ & 0.013 -- 0.05 & 0.01 -- 0.06 \\
$ \sin^2\theta_{24} $  & 0.023$_{-0.008}^{+0.004}$ & 0.005 -- 0.035 & 0.003 -- 0.04 \\
$ \sin^2\theta_{34} $ &  --  & -- &$ <$ 0.18  \\
\hline
\end{tabular}
\begin{center}
\caption{Best-fit and $3 \sigma$ ranges of active neutrino oscillation parameters as given in \cite{Gonzalez-Garcia:2015qrr,Forero:2014bxa,Capozzi:2013csa}.
The current constraints on sterile neutrino parameters are from the global analysis as
given in \cite{Giunti:2011gz,schwetz}. Here, $ \Delta m_{LSND}^2$ $ (\mathrm{eV}^2) $ is either  $\Delta m_{41}^2 (\mathrm{eV}^2)$ or $\Delta m_{43}^2 (\mathrm{eV}^2)$.}
\label{parameters}
\end{center}
\end{table}
In the 3+1 scenario, the 4$\times$4 Majorana neutrino mass matrix, $M_{\nu}$, in the flavor basis
 can be diagonalized as
 \be
 M_{\nu} = V M_{\nu}^{diag}V^T,
\ee
where $M_{\nu}^{diag}$=diag$\{m_1,m_2,m_3,m_4\}$ is the diagonal neutrino mass matrix. 
The matrix $V$ is the leptonic mixing matrix in the basis where the charged lepton mass matrix is diagonal \cite{Goswami:2005ng} and is given as
\be
\label{U}
\rm V = U.P.
\ee
In general, any arbitrary $N \times N$ unitary matrix contains $\frac{N(N-1)}{2}$ mixing angles
and $\frac{1}{2}(N-1)(N-2)$ Dirac type CP phases. If the neutrinos are Majorana particles
there will be additional ($N-1$) Majorana phases.
Here $U$ is the lepton mixing matrix for Dirac neutrinos and can be written as
\begin{equation}
U={R_{34}}\tilde R_{24}\tilde R_{14}R_{23}\tilde R_{13}R_{12},
\end{equation}
where $R_{ij}$ denotes rotation matrices in the \textit{ij} generation space
and is expressed as, 
\begin{center}
$R_{23}$=$\left(
\begin{array}{cccc}
1~ &~0 & 0 & 0 \\  0~ &~ c_{23} & s_{23} & 0 \\ 0~ & -s_{23} & c_{23}& 0 \\0 ~& ~0 & 0 & 1
\end{array}
\right)$ , $\tilde {R}_{13}$=$\left(
\begin{array}{cccc}
c_{13}~ & ~0 &~ s_{13}e^{-i \delta_{13}} &~ 0\\ 0 ~ & ~ 1&~~ 0 & 0 \\ ~ -s_{13}e^{i \delta_{13}}
 ~& ~0 &~~ c_{13} & 0 \\ 0  & ~ 0& ~~0 & 1
\end{array}
\right),$ \\
\end{center}
where $s_{ij}=\sin\theta_{ij}$ and $c_{ij}=\cos\theta_{ij}$. The mixing
 matrix $U$ is parameterized in terms of six mixing angles: $\theta_{12}$, 
$\theta_{23}$, $\theta_{13}$, $\theta_{14}$, $\theta_{24}$ and
 $\theta_{34}$ 
 and three Dirac CP violating phases: $\delta_{13}$, $\delta_{14}$, $\delta_{24}$. 
 The diagonal phase matrix $P$ in Eq (\ref{U}) is given as, 
\begin{equation}
P=diag\{1,e^{i\alpha},e^{i(\beta+\delta_{13})}, e^{i(\gamma + \delta_{14})}\},
\end{equation}
where $\alpha$, $\beta$ and $\gamma$ are three Majorana CP violating phases.
Table. \ref{parameters}, gives the current best-fits and $3\sigma$ ranges of the mixing parameters.
Oscillation parameters for active neutrinos are taken from \cite{Gonzalez-Garcia:2015qrr,Forero:2014bxa,Capozzi:2013csa} and
sterile mixing parameters are taken from \cite{Kopp:2013vaa,schwetz}.

Oscillation experiments can measure only the mass squared differences and give no information
on the absolute neutrino masses. Information on absolute masses can come from neutrinoless double 
beta decay ($0\nu \beta\beta$) experiments. Double beta decay experiments will observe the lepton number violating process
and probe the Majorana nature of neutrinos. Neutrinoless double beta decay processes are 
sensitive to effective Majorana mass which in 3+1 scenario is given as

\bea
\label{effectivemm}
M_{ee} &=& \nn |\Sigma U_{ei}^2 m_i| \\ 
&=& |(U_{e1})^2 m_1 + (U_{e2})^2 m_2 e^{2i\alpha} \\ \nonumber
&+&(U_{e3})^2 m_3 e^{2i(\beta+\delta_{13})}+(U_{e4})^2 m_4 e^{2i(\gamma+\delta_{14})}| \\ \nonumber
&=& | m_1 c_{12}^2c_{13}^2c_{14}^2 +m_2 e^{2i\alpha}c_{13}^2c_{14}^2s_{12}^2 \\ \nonumber 
&+& m_3 e^{2i\beta}s_{13}^2c_{14}^2 + m_4 e^{2i\gamma}s_{14}^2| .
\eea
We see that $M_{ee}$ is sensitive to the Majorana phases which are
 present in the neutrino mass matrix. A large number of projects such as
 CUORE \cite{Gorla:2012gd}, GERDA \cite{Wilkerson:2012ga}, SuperNEMO \cite{Barabash:2012gc}, KamLAND-ZEN \cite{Gando:2012zm} and  EXO \cite{Auger:2012ar} aim to 
 discover evidences for neutrinoless double beta decay. These experiments also put bounds on the effective Majorana mass $M_{ee}$. These experiments are expected to
 achieve a sensitivity upto 0.01 eV for $M_{ee}$.
The combined results from KamLAND-ZEN and
EXO-200 \cite{Gando:2012zm} give the upper bound on the effective Majorana neutrino mass as $M_{ee}<$ (0.12 - 0.25) eV, 
where the range comes from uncertainty in the nuclear matrix elements.
The next generation experiments are expected to improve the bounds and the sensitivity to $M_{ee} <  (0.0080 \pm 0.0016)$ eV \cite{DellOro:2014yca}. 

\section{Formalism}
\label{sec3}

One-zero texture neutrino mass matrix results in the following condition
\bea
\label{texturezero}
&&m_1 U_{a1}U_{b1} + m_2 U_{a2}U_{b2} e^{2i\alpha} \\ \nonumber 
&+& m_3 U_{a3} U_{b3} e^{2i(\beta + \delta_{13})}
+ m_4 U_{a4} U_{b4} e^{2i(\gamma + \delta_{14})} = 0
\eea
where $a, b$ can be $e$, $\mu$, $\tau$ and $s$. 
We study the zero texture in 4$\times$4 neutrino
mass matrix taking: (i) vanishing lowest mass i.e., $m_1 = 0$ for NH (A) and $m_3 = 0$ for IH (B) and (ii) the three active masses are almost 
equal i.e., $m_1 \sim m_2 \sim m_3$ (C).

\subsection{Normal Hierarchy}

For normal hierarchy, one textures with a vanishing lowest mass ($ m_{1} = 0 $), Eq. (\ref{texturezero}) can be written as,
\begin{equation}
p U_{a2}U_{b2} + U_{a3} U_{b3} e^{2i(\beta + \delta_{13})} + q U_{a4} U_{b4}  e^{2i(\gamma + \delta_{14})} = 0. 
\label{nh}
\end{equation}
Here, $p= \frac{m_2}{m_3}e^{2i\alpha}$ and $q = \frac{m_4}{m_3}$. \\
Solving the above equation constrains $p$ as
\be
p = -\left(\frac{U_{a3}U_{b3}e^{2i(\beta + \delta_{13})} + q U_{a4}U_{b4}e^{2i(\gamma + \delta_{14})}}{U_{a2}U_{b2}}\right).
\ee
The neutrino  mass ratio for normal hierarchy is
\bea 
\frac{m_2}{m_3} &=& \nonumber |p|
\eea
and one of the physical Majorana phase $\alpha$ is given as
\begin{equation}
\alpha = \frac{1}{2}arg (p).
\end{equation}
The total number of free parameters is five, three Dirac and two Majorana CP
phases. Note that we have assumed one of the active neutrino mass to vanish so
the number of physical phases is two. For example we can always take the phase $\beta$ out of the Eq. (\ref{nh}) and
the physical Majorana phases will be $(\alpha-\beta)$ and $(\gamma - \beta)$.
The condition of vanishing lowest mass ensures that all the remaining three
masses are well determined from the experimental mass squared difference.
The remaining three masses for normal hierarchy are found in terms of mass squared differences as
$m_2 = \sqrt{\Delta m_{21}^2}$, $m_3 = \sqrt{\Delta m_{21}^2 +\Delta m_{31}^2}$ and $m_4 =
\sqrt{\Delta m_{41}^2}$.
We span the parameter space of input neutrino oscillation parameters
(3 active and 3 active sterile mixing angles) lying in their 3$\sigma$ ranges by randomly
generating points to the order of 10$^9$. Since the Dirac CP phases are experimentally
unconstrained at 3$\sigma$ level, therefore, we vary them within their full possible range
[0$^\circ$ -- 360$^\circ$]. We can find two independent mass squared difference ratios defined as,
\bea \rm
R_{\nu} &=& \nonumber \frac{\Delta m_{21}^2}{\Delta m_{32}^2} = \frac{|p|^2}{1-|p|^2}, \\
 \rm R_{\nu_1} &=& \frac{\Delta m_{41}^2}{\Delta m_{32}^2} = \frac{q^2}{1-|p|^2}.
\eea
The allowed 3$\sigma$ ranges for these mass squared difference ratios is
\bea
\rm R_{\nu} &=& \nn (2.7 \times 10^{-2} - 3.5 \times 10^{-2}), \\
\rm R_{\nu_1} &=&  (0.26 \times 10^{3} - 1.08 \times 10^{3}).
\eea
The given texture is viable if the mass squared difference ratios are in their current 3$
\sigma$ ranges.
Since we consider the neutrino mass matrices with lowest
vanishing mass ($ m_{1} = 0 $ for NH), in that case using the fact that  $R_{\nu} \ll 1$, the masses in terms of $R_{\nu}$ and $R_{\nu1}$ can be rewritten as,
\begin{widetext}
\bea
\label{xnh}
\rm SNH: |m_4| \approx \sqrt{ \Delta m_{31}^2 R_{\nu1}} \gg |m_3|
\approx \sqrt{(1 + R_{\nu})\Delta m_{31}^2} \approx \sqrt{\Delta m_{31}^2}
\gg |m_2| \approx \sqrt{\Delta m_{31}^2 R_{\nu}}.
\eea
\end{widetext}

\subsection{Inverted Hierarchy}
For inverted hierarchy with mass $m_3 = 0$, Eq. (\ref{texturezero}) can be written as
\begin{equation}
U_{a1}U_{b1} + p U_{a2}U_{b2} + q U_{a4} U_{b4} e^{ 2i \delta_{14}}= 0,
\label{ih}
\end{equation}  
where $p= \frac{m_2}{m_1}e^{2i\alpha}$ and $q = \frac{m_4}{m_1}e^{2i\gamma}$. Solving Eq.
(\ref{ih}) gives
\be
p = -\left(\frac{U_{a1}U_{b1} + q U_{a4}U_{b4}e^{2i\delta_{14}}}{U_{a2}U_{b2}}\right).
\ee
We can extract the neutrino mass ratio as
\bea
\frac{m_2}{m_1} &=& \nonumber |p|
\eea
and one of the Majorana phases is found to be
\begin{equation}
\alpha = \frac{1}{2}arg (p).
\end{equation}

Here the physical Majorana phases are $\alpha$ and $\gamma$.
For inverted hierarchy the remaining three neutrino masses are given as 
$m_1 = \sqrt{\Delta m_{31}^2}$, $m_2 = \sqrt{\Delta m_{21}^2 +\Delta m_{31}^2}$ and $m_4 =
\sqrt{\Delta m_{43}^2}$. We define the two independent neutrino mass squared difference ratios as,
\bea \rm
R_{\nu} &=& \nonumber \frac{\Delta m_{21}^2}{\Delta m_{31}^2} = |p|^2 -1, \\
\rm R_{\nu_1} &=& \frac{\Delta m_{43}^2}{\Delta m_{31}^2} = |q|^2,
\eea
where $|q| = \frac{m_4}{m_1}$.
The allowed 3$\sigma$ ranges of these two mass ratios calculated from the experimental data are 
\bea
R_{\nu} &=& \nn (2.7 \times 10^{-2} - 3.7 \times 10^{-2}), \\
R_{\nu_1} &=&  (0.27 \times 10^{3} - 1.14 \times 10^{3})
\eea
For the case of IH with $m_3=0$ and using $R_{\nu} \ll 1$ the masses can be written in terms of 
$R_{\nu_1}$ as
\begin{widetext}
\bea \rm
\label{xih}
SIH: |m_4|\approx \sqrt{\Delta m_{31}^2 R_{\nu_1}} \gg|m_2|
\approx \sqrt{(1 + R_{\nu})\Delta m_{31}^2} \approx \sqrt{\Delta m_{31}^2}
\approx |m_1| \approx \sqrt{\Delta m_{31}^2}.
\eea
\end{widetext}
\subsection{Quasidegenerate Mass Spectrum}
 For quasidegenerate mass spectrum the three active neutrinos have approximately
same mass and are lighter compared to the mass of sterile neutrino, 
\begin{equation}
QD : |m_{4}| >> |m_{1}| \approx |m_{2}| \approx |m_{3}| \approx m_{0}.
\label{qd}
\end{equation} 
Using QD approximation the one-zero texture equation (Eq. (\ref{texturezero})) can be written as,
\bea
\label{texturezero_qd}
&&m_0( U_{a1}U_{b1} +  U_{a2}U_{b2} e^{2i\alpha} 
+  U_{a3} U_{b3} e^{2i(\beta + \delta_{13})} )  \\ \nonumber
&+&  m_4 U_{a4} U_{b4} e^{2i(\gamma + \delta_{14})}  = 0
\eea
In the QD regime we vary $m_0$ from 0.1 eV - 0.3 eV.
\section{Results}
\label{sec4}

In this section we first study the implications of a vanishing element
$M_{\alpha \beta}$ with vanishing lowest mass for the 3+1 scenario, where $\alpha, \beta = e, \mu,\tau,
 s$ and then study the implications of quasidegenerate neutrino mass spectrum in the active sterile mixing. 
The vanishing lowest neutrino mass in
addition to the zero texture have non-trivial implications and results
in the constrained parameter space of neutrino masses and CP violating phases. 
Interesting parameters correlations are also obtained for the quasidegenerate case.
 Since $M_{\alpha \beta}$ is complex the above condition implies 
both real and imaginary parts are zero. Therefore to study the one-zero textures we consider $|M_{\alpha \beta}|=0$.

\subsection {The mass matrix element $M_{ee}$}
\begin{figure*}
\begin{center}
\includegraphics[width=0.47\textwidth]{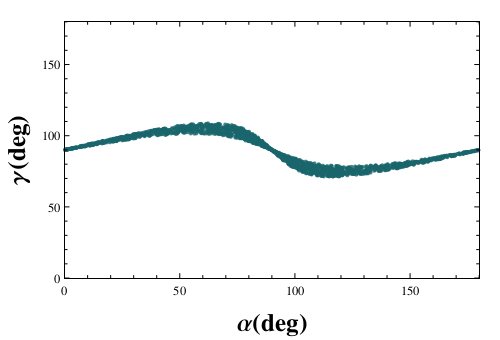}
\includegraphics[width=0.47\textwidth]{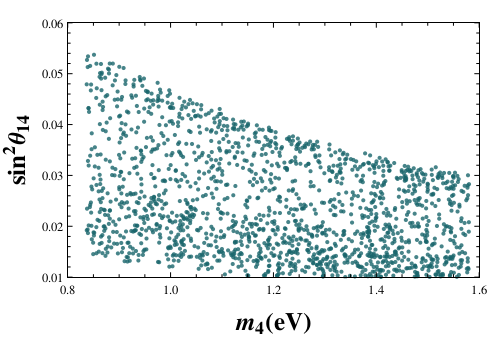}\\
\includegraphics[width=0.47\textwidth]{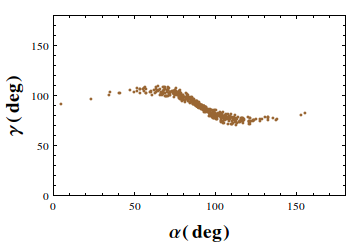}
\includegraphics[width=0.47\textwidth]{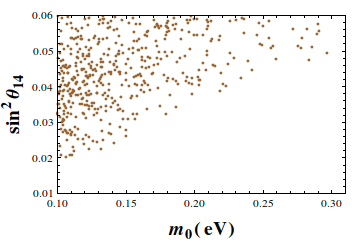}
\caption{Correlation plots for $|M_{ee}|$ = 0, IH (upper panel) and quasidegeneracy (lower panel).}
\label{figmeeih}
\end{center}
\end{figure*}
%
The matrix element $M_{ee}$ in the 3+1 scenario is given as,
\bea
M_{ee}&=&m_1 c_{14}^2 c_{13}^2c_{12}^2+m_2 s_{12}^2c_{14}^2c_{13}^2e^{2i\alpha} \\ \nonumber
&+&m_3 s_{13}^2 c_{14}^2 e^{2i\beta}+m_4s_{14}^2 e^{2i\gamma}.
\label{mee} 
\eea
The 3+1 picture have predictions that are completely different from standard 3 active scenario. $M_{ee}$ cannot vanish in 3 active scenario if the lowest mass vanishes \cite{Merle:2006du,Lashin:2011dn}.
In 3+1 case we find that it can vanish due to the contribution from the sterile sector.
The  contribution of the sterile neutrino to the element $M_{ee}$ comes from the mass $m_4$ and the active-sterile mixing angle $\theta_{14}$. The mass matrix element $M_{ee}$ has the simplest form because of the chosen parametrization and can be understood quite well. \\
For NH ($m_1=0$) the zero-texture at $M_{ee}$ can be written as,
\be 
\label{meenh}
|c_{14}^2(\sqrt{R_{\nu}}s_{12}^2c_{13}^2e^{2i\alpha}+s_{13}^2 
e^{2i\beta} )+ \sqrt{R_{\nu_1}}s_{14}^2 e^{2i\gamma}| = 0.
\ee
For $\theta_{14} = 0^\circ$ this expression reduces to one-zero texture in 3 active neutrino mixing scenario.
In the three active neutrino case for $\beta = 0^\circ$ and $\alpha = +90^\circ$ (which maximizes the cancellation between the terms) $M_{ee}$ predicts
\begin{equation}
 s_{13}^2 = \sqrt{R_\nu} s_{12}^2 c_{13}^2,
\end{equation}
which gives $s_{13}^2 = 0.047$ for the lowest value of $R_\nu$ and this is higher than the experimental prediction. Thus for three active scenario $M_{ee}$ can not be zero for NH when 
$m_1$ is zero \cite{Gautam:2015kya}. We found earlier in our 3+1 scenario analysis \cite{Ghosh:2013nya} that the cancellation of terms on the RHS of Eq. (\ref{mee}) is possible only if $m_1$ is large. In our present case since lowest mass is  vanishing, the major contribution will come from additional sterile neutrino because of  higher values of $m_4$. Even for very small allowed values of $\theta_{14}$ the sterile
contribution (third term) in Eq. (\ref{meenh}) is large  and cannot have equal
 magnitude to the other two terms on its left. Thus, complete cancellation of the terms is
never possible resulting in non vanishing $M_{ee}$ for NH. The only possible way to have
vanishing $M_{ee}$ is when $\theta_{14}$ is very small. It's typical value
 can be obtained from Eq. (\ref{meenh}) as,
\be 
\rm tan^2 \theta_{14} = - \frac{s_{13}^2e^{2i\beta}+\sqrt{R_{\nu}}s_{12}^2c_{13}^2e^{2i\alpha}}{\sqrt{R_{\nu_1}}e^{2i\gamma}}.
\ee
Substituting $\alpha = \beta = 0^\circ$ (which would maximize the active neutrino contribution) and 
$\gamma = +90^\circ$ we get $ \tan^2 \theta_{14} \approx 10^{-3}$ i.e., $\theta_{14}$ lies
well outside its allowed 3$\sigma$ range and is thus not allowed.\\
For IH when $m_3 = 0$, zero texture condition for $M_{ee}$ can be written as,
\be 
\label{meeih}
 |c_{14}^2c_{13}^2(c_{12}^2 + s_{12}^2 e^{2i\alpha})+
\sqrt{R_{\nu_1}}s_{14}^2 e^{2i\gamma} |=0. 
\ee
As can be seen from above equation the magnitude of $M_{ee}$ for IH depends on $\alpha$ and $\gamma$ along with the mixing angles. For IH the complete cancellation is never possible in three active neutrino scenario because in addition to $\alpha = +90^\circ$ cancellation requires $s_{12}^2 = c_{12}^2$ \cite{Merle:2006du, Lashin:2011dn}. These results change when we include the sterile contribution. The element $M_{ee}$ vanishes for IH in the limit $m_3\approx 0$  when $\alpha = 0^\circ$  and $\gamma = +90^\circ$ provided 
\be 
\tan^2\theta_{14}  \approx \frac{c_{13}^2}{\sqrt{R_{\nu_1}}} \approx 0.05
\ee
which is well within the allowed range. This behaviour is in stark contrast to that of the 3 neutrino 
case \cite{Merle:2006du}. Note that, by considering vanishing lowest mass, the other two masses are highly constrained in the present scenario. $m_1 = \sqrt{\Delta m_{31}^2}$ and $m_2=\sqrt{\Delta m_{31}^2 + \Delta m_{21}^2}$ for IH. From Eq. (\ref{meeih}) it is evident that the cancellation of all three terms are dependent on the two Majorana phases $\alpha$ and $\gamma$. The order of magnitude of each term is approximately the same of the order of $\mathcal{O}$(10$^{-2}$). The cancellation of these terms are only possible for constrained values of Majorana phase $\gamma$ (around $+90^\circ$) as shown in the upper left panel of Fig. (\ref{figmeeih}). Note that as $m_4$ is large compared to $m_1$ and $m_2$, for cancellation between the active and sterile sector, one expects to have lower values of $s_{14}^2$ for high $m_4$ values.
This is evident from the upper right panel of Fig. (\ref{figmeeih}) where we can see that
the higher values of $\theta_{14}$ are disallowed as the value of $m_4$ increases.\\
The matrix element $M_{ee}$ for QD neutrino masses is given as,
\bea \label{meeqd} 
M_{ee}  &= &m_0 c_{14}^2 c_{13}^2 (c_{12}^2 + s_{12}^2 e^{2i\alpha})  \\ \nonumber
&& + m_0 s_{13}^2 c_{14}^2 e^{2i\beta} + m_4s_{14}^2 e^{2i\gamma}.
\eea
The correlation plots for QD mass spectrum when $M_{ee}$ vanishes are given in the lower panels of Fig. (\ref{figmeeih}). We observe that unlike the completely hierarchical mass spectrum (upper left panel of Fig. (\ref{figmeeih})), here, the phase $\alpha$ is also constrained. For $\alpha = 0^\circ$ or 180$^\circ$ and $\beta = 0^\circ$ the above expression reduces to 
\bea
M_{ee}&=&m_0 c_{14}^2 + m_4 s_{14}^2 e^{2i \gamma}.
\eea
When $\gamma$ $\sim$ 90$^\circ$ the term $m_0 c_{14}^2$ becomes order of magnitude greater than
the second term and hence the cancellation of these two terms becomes impossible. 
We see this behavior in the ($\alpha - \gamma$) plot in the lower left panel of Fig. (\ref{figmeeih}) 
where $\alpha = 0^\circ$ or 180$^\circ$ is disallowed. However, when $\alpha = \beta= \gamma$ = 90$^\circ$, Eq. (\ref{meeqd}) becomes
\be 
M_{ee} = m_0 c_{14}^2 (c_{13}^2 \cos 2\theta_{12} - s_{13}^2) - m_4 s_{14}^2.
\label{mee2}
\ee
Now all terms in the above expression are of same order. Thus $M_{ee}$ can vanish for these values
of Majorana phases $\alpha$ and $\gamma$. Note that the other phase $\beta$ is unconstrained
here.
From the above expressions it is clear that vanishing of $M_{ee}$ for QD mass spectrum depends on
active sterile mixing angle $\theta_{14}$ along with the Majorana phases $\alpha$ and $\gamma$. In the lower right panel of Fig. (\ref{figmeeih}), we plot $\sin^2\theta_{14}$ as a function of $m_0$. 
As $m_0$ increases, $\theta_{14}$ constrains to higher values. This behaviour can
also be understood from Eq (\ref{mee2}). As the active neutrino mass
increases, higher values of $\theta_{14}$ are needed to allow
cancellation between the active and the sterile terms. 

\begin{figure*}
\begin{center}
\includegraphics[width=0.47\textwidth]{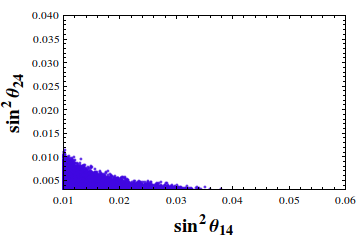}
\includegraphics[width=0.47\textwidth]{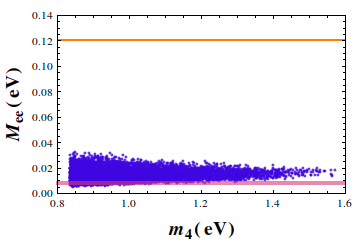}\\
\includegraphics[width=0.47\textwidth]{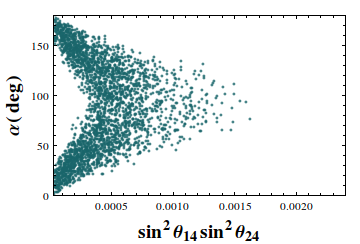}
\includegraphics[width=0.47\textwidth]{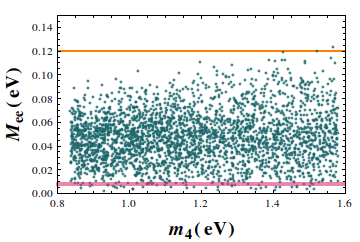}\\
\includegraphics[width=0.47\textwidth]{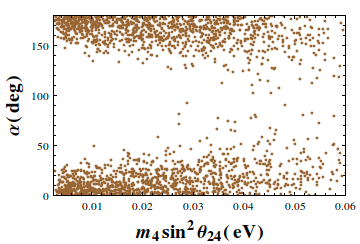}
\includegraphics[width=0.47\textwidth]{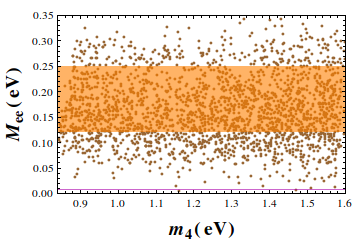}
\end{center}
\begin{center}
\caption{Correlation plots for $|M_{e\mu}|$ = 0. Upper (middle) row is for NH (IH) and lower row is for QD mass spectrum. Orange line in the upper and middle right panels represent the lowest value of the upper bound on $M_{ee}$ from KamLAND-ZEN $ + $ EXO-200. The purple line is the sensitivity from next generation experiments \cite{DellOro:2014yca}. The orange band in the lower right panel represents the full allowed range of $M_{ee}$ from combined results of KamLAND-ZEN and EXO-200.}
\label{figmemuih}
\end{center}
\end{figure*}


\subsection {The mass matrix elements $M_{e\mu}$ and $M_{e\tau}$ }

The mass matrix element $M_{e\mu}$ in the 3+1 scenario is given as,
\begin{widetext}
\bea
M_{e \mu}&=& m_4 c_{14}(e^{i (\delta _{14}-\delta _{24}+2 \gamma)}s_{14}s_{24}
+m_3 e^{i(\delta_{13}+2 \beta)} s_{13}(c_{13}c_{24}s_{23}-e^{i
(\delta_{14}-\delta_{13}-\delta_{24})} s_{13}s_{14}s_{24}) \\ \nonumber
&+&m_1 c_{12}c_{13}(-c_{23}c_{24}s_{12}
+c_{12}(-e^{i \delta_{13}}c_{24}s_{13}s_{23}-e^{i (\delta_{14}-\delta_{24})}c_{13}s_{14}s_{24})) \\
\nonumber &+&m_2 e^{2i \alpha}c_{13}s_{12} (c_{12}c_{23}c_{24}
+s_{12}(-e^{i \delta_{13}}c_{24}s_{13}s_{23}-e^{i (\delta_{14}-\delta_{24})}c_{13}s_{14}s_{24}))).
\label{me}
\eea
\end{widetext}
This expression is complicated as compared to $M_{ee}$ and thus it becomes impossible to study the behaviour of independent parameters. 
To simplify this expressions we introduce a small parameter $\lambda \equiv$ 0.2 and retain
terms of the order of $\lambda^2$.
We define the small angles $\theta_{14}$, $\theta_{24}$ and $\theta_{13}$ in the form a$\lambda$.
 \bea \nonumber \label{a1}
{\mathrm{sin}} \theta_{14} \approx \theta_{14} \equiv a_{14}\lambda, \\
{\mathrm{sin}} \nonumber \theta_{24} \approx \theta_{24} \equiv a_{24}\lambda,\\
\sin \theta_{13} \approx \theta_{13} \equiv a_{13}\lambda,
\eea

where $a_{ij}$ are parameters of $\mathcal{O}$(1) and their $3 \sigma$ range
from the current constraint on the mixing angles is given by
\begin{eqnarray}
a_{13} &=& 0.68 - 0.81, \\ \nonumber
a_{14} &=& 0.5 - 1.2, \\ \nonumber
a_{24} &=& 0.25 - 1.
\end{eqnarray}
Note that the sterile mixing angle $\theta_{34}$ can be large (has an upper bound) and thus we do not approximate
this angle. We will use the above approximations to understand the behaviour of different parameters wherever required.

The condition of zero texture at $M_{e\mu}$ for NH with vanishing lowest mass ($ m_{1} = 0 $) gives,
\bea
 && |m_4 c_{14}(e^{i (\delta _{14}-\delta _{24}+2 \gamma)}s_{14}s_{24} \\ \nonumber
&+& m_3 e^{i(\delta_{13}+2 \beta)} s_{13}(c_{13}c_{24}s_{23}-e^{i
(\delta_{14}-\delta_{13}-\delta_{24})} s_{13}s_{14}s_{24})\\ \nonumber
&+& m_2e^{2i \alpha}c_{13}s_{12} (c_{12}c_{23}c_{24} \\ \nonumber
&+&s_{12}(-e^{i \delta_{13}}c_{24}s_{13}s_{23}-e^{i (\delta_{14}-\delta_{24})}c_{13}s_{14}s_{24})))| = 0.
\label{me}
\eea
From the above expression we see that the sterile sector contribution appears as $\theta_{14}$ and $\theta_{24}$. For cancellation of sterile term with the active part, 
the values of these angles must be small. This is clear from the upper left panel of Fig. (\ref{figmemuih}) where we give the correlation
of $s_{14}^2$ with $s_{24}^2$. We see that the higher values
of $s_{14}^2$ and $s_{24}^2$ are disallowed and as $s_{14}^2$ increases, $s_{24}^2$ decreases.
In the upper right panel of Fig. (\ref{figmemuih}) we give the prediction for the effective mass $M_{ee}$ as a function of $m_4$.
As $M_{ee}$ is proportional to $s_{14}^2 m_4$ it is expected that as $m_4$ increases, one should get higher values of effective Majorana mass.  
But in this particular case we 
do not see this feature. This is because for cancellation to occur between the active and sterile terms in $M_{e \mu}$,
one needs smaller $\theta_{14}$ for the higher values 
of $m_4$ and thus higher values of $M_{ee}$ are not possible even when $m_4$ is large.
Thus, for NH and $|M_{e\mu}| = 0$ there exists an upper bound on effective mass $M_{ee} < 0.04$ eV.\\
The zero texture at $M_{e\mu}$ for IH with vanishing lowest mass ($ m_{3} = 0 $) in 3+1 scenario using approximations given in Eqs. (\ref{xih}, \ref{a1}) can be written as,
\bea  \label{memih}
 && |c_{12}s_{12}c_{23}(e^{2i\alpha}-1) \\ \nonumber
&-& s_{23}a_{13}e^{i\delta_{13}}(c_{12}^2+s_{12}^2 e^{2i\alpha})\lambda \\ \nonumber
&-& e^{i(\delta_{14}-\delta_{24})}
a_{14}a_{24}(c_{12}^2 -e^{2i\gamma}\sqrt{R_{\nu_1}} +e^{2i\alpha}s_{12}^2)\lambda^2 |=0,
\eea
which is independent of $\theta_{34}$ and Majorana phase $\beta$. 

From the above equation it can be noted that for $\alpha = 0^\circ$, the leading order term vanishes. Thus for cancellation to occur one needs very small values of $\theta_{14}$ and $\theta_{24}$ which appears with the large $m_4$ term. On the other hand when $\alpha$ is $+90^\circ$, the leading order term is 
very large and one requires higher values of the above mentioned mixing angles for cancellation. This is shown in the middle left panel of Fig. (\ref{figmemuih}). 
Here, for $\alpha = 0^\circ$, the allowed values of $\sin^2\theta_{14} \sin^2\theta_{24}$ are very
small and on the contrary, for $\alpha = +90^\circ$ the smaller values of  $\sin^2\theta_{14} \sin^2\theta_{24}$ are disallowed. 
Here the upper bound of effective mass $M_{ee}$ can be as high as $0.13$ eV (middle right panel of Fig. (\ref{figmemuih})).

In these figures (as well as for the other allowed textures) we also show the present lowest value of the upper bound on $M_{ee}$ by 
the orange line and the future expected sensitivity by the purple line. 
These figures clearly show that the values of $M_{ee}$ are always lower than the present bound on $M_{ee}$ for NH. 
Whereas for IH, values of $M_{ee}$ are below the current bound for smaller value of $ m_{4} $ but
for higher value of  $ m_{4} $ there are a few points of $M_{ee}$ which are above the lowest value of the current upper bound i.e., above the orange line. 
The above statement is also true for all the $M_{ee}$ figures for completely hierarchical mass presented in the subsequent sections. \\
Using approximations given in Eq.(\ref{qd}, \ref{a1}) for quasidegeneracy, the zero texture at $M_{e\mu}$ for QD spectrum can be written as,
\bea  \label{memqd}
 && |m_0 [c_{12}s_{12}c_{23}(e^{2i\alpha}-1) \\ \nonumber
&+&a_{13} s_{23}e^{i\delta_{13}} 
(- c_{12}^2+e^{2 i\beta}-s_{12}^2 e^{2i\alpha})\lambda] \\ \nonumber
&& - a_{14}a_{24} e^{i\delta_{14}} (c^{2}_{12}m_{0} - e^{2 i\gamma} m_{4}+  e^{2 i\alpha}  s^{2}_{12}m_0)\lambda^2|=0. \\ \nonumber
\eea
The correlation plots for the parameters here are given in the lower panels of Fig. (\ref{figmemuih}). The Majorana phase $ \alpha $ is largely constrained
(values near 0$^\circ$ or 180$^\circ$ are seems to be allowed) for smaller values of $m_4\sin^{2}\theta_{24} $. For larger values of $ m_4\sin^{2}\theta_{24}$ this texture is marginally allowed. To understand this behaviour, we consider $ \alpha = 90^\circ $ and all other phases to be $0^\circ$
and express Eq. (\ref{memqd}) as,
\bea \label{memqd1}
&& |m_0[-2 (c_{12} c_{23} s_{12}) + 2 a_{13}s^2_{12}s_{23} \lambda] \\ \nonumber
&+&(a_{14}a_{24}[m_{4}-\cos2\theta_{12}m_0])\lambda^{2}|=0.
\eea
From the equation it is clear that, the $m_4 a_{24}$ term with a coefficient $\lambda^2$ (which is small), can only be of the similar order with the leading order $m_0$ term, when $m_4$ and $\theta_{24}$ is large.
Thus for smaller values of
$m_4 s_{24}^2$ it is not possible to have any cancellation and thus there are no viable points around $\alpha$  = 90$^\circ$
as seen in the lower left panel of Fig. (\ref{figmemuih}).


However, for $ \alpha = 0^\circ $ or $ 180^\circ $, the leading order $m_0$ term and the subleading term
with $\lambda$ can vanish due to the
presence of $ (e^{2i\alpha}-1)$ term. For this reason one can always have cancellation within the remaining $\lambda^2$ terms,
for all the values of $m_4$ and $s_{24}$ (lower left panel of Fig. (\ref{figmemuih})).

In the lower right panel of Fig. (\ref{figmemuih}), we plot the effective Majorana mass $M_{ee}$ vs $m_4$.
The orange band in this plot, corresponds to the allowed range of $M_{ee}$ with the nuclear matrix element uncertainty as given by the combined analysis of 
KamLAND-ZEN and EXO-200 (Xe$^{136} $). 
From this plot we observe that a certain region of the allowed parameter space can be completely
discarded based on the present bound on $M_{ee}$. However, this is not the case for completely hierarchical neutrino masses. For NH, the allowed values are well below of the current upper limits of $M_{ee}$
and for IH, there are few points which fall in the region of uncertainty i.e., points above the orange line.
The conclusion drawn in this section is also true for the next subsequent sections and that can be seen from the right panel figure of each allowed texture.

\begin{figure*}
\begin{center}
\includegraphics[width=0.47\textwidth]{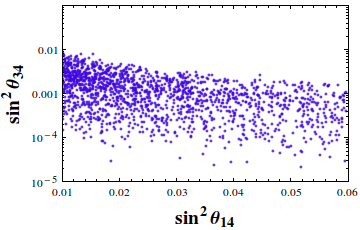}
\includegraphics[width=0.47\textwidth]{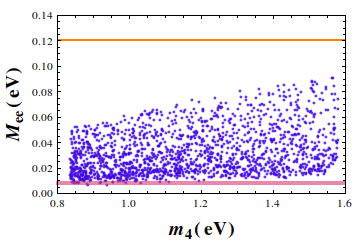}\\
\includegraphics[width=0.47\textwidth]{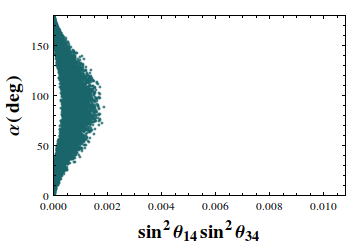}
\includegraphics[width=0.47\textwidth]{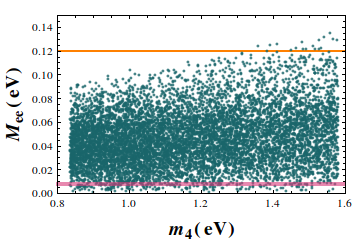}\\
\includegraphics[width=0.47\textwidth]{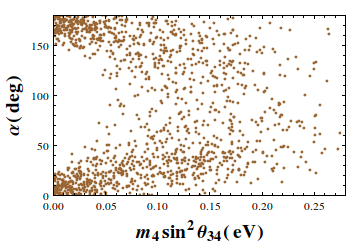}
\includegraphics[width=0.47\textwidth]{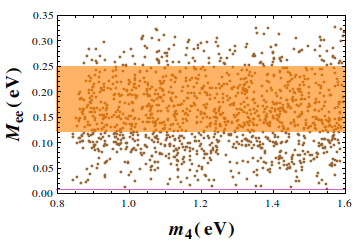}\\
\end{center}
\begin{center}
\caption{Correlation plots for $|M_{e\tau}|$ = 0 for NH (upper panel), IH (middle panel) and QD (lower panel) mass spectrum. Bounds on $M_{ee}$ are described in the caption of Fig.(\ref{figmemuih}).}
\label{figmetau}
\end{center}
\end{figure*}

The mass matrix element $M_{e\tau}$ in the 3+1 mixing scenario is given as,
\bea
M_{e\tau}&=&c_{14}c_{24}e^{i(2\gamma + \delta_{14})}m_4s_{14}s_{34} \\ \nonumber
&+&m_3c_{14}s_{13}e^{i(2\beta+\delta_{13})}(-c_{24}s_{13}s_{14}s_{34}
e^{i(\delta_{14}-\delta_{13})}\\ \nonumber
&+&c_{13}(c_{23}c_{34}-e^{i\delta_{24}}s_{23}s_{24}s_{34})) \\ \nonumber
&+&m_2s_{12}c_{13}c_{14}e^{2i\alpha}(c_{12}(-c_{34}s_{23}-c_{23}s_{24}s_{34}e^{i\delta_{24}}) \\ \nonumber
&+&s_{12}(-c_{13}c_{24}s_{14}s_{34}e^{i\delta_{14}} \\ \nonumber
&-&e^{i\delta_{13}}s_{13}(c_{23}c_{34}-e^{i\delta_{24}}s_{23}s_{24}s_{34}))) \\ \nonumber
&+&m_1c_{12}c_{13}c_{14}(-s_{12}(-c_{34}s_{23}-c_{23}s_{24}s_{34}e^{i\delta_{24}}) \\ \nonumber
&+& c_{12}(-c_{13}c_{24}s_{14}s_{34}e^{i\delta_{14}}
-e^{i\delta_{13}}s_{13}\\ \nonumber
&&(c_{23}c_{34}-e^{i\delta_{24}}s_{23}s_{24}s_{34}))).
\eea
The two elements $M_{e\mu}$ and $M_{e\tau}$ in 3 active neutrino scenario are related by 
$\mu-\tau$ permutation symmetry \cite{Lam:2005va,Grimus:2001uc,Mohapatra:2005yu,Gupta:2013it}. Here in 3+1 case also these two elements are 
related as \begin{center}$(M_\nu)_{e\tau}=P_{\mu\tau}^T (M_\nu)_{e\mu} P_{\mu\tau} . $ \end{center} where permutation matrix $P_{\mu \tau}$ is given as
\begin{center}
$
P_{\mu\tau}=\left(
\begin{array}{cccc}
 1& 0 & 0 & 0 \\ 0& 0 &1& 0 \\ 0& 1 &0 & 0\\ 0 & 0 & 0 &1
\end{array}
\right)$.
\end{center}
For three active neutrino case the atmospheric mixing angle $\theta_{23}$ 
in the $\mu-\tau$ symmetric textures are related as
$\bar\theta_{23}=(\frac{\pi}{2} -\theta_{23})$.
However, in the 3+1 case the relation of $\theta_{23}$ between two texture structures related by this symmetry is not simple \cite{Ghosh:2012pw}. The active sterile mixing angles $\theta_{24}$ and $\theta_{34}$ are also different and are related as,
\begin{equation}
\bar\theta_{12}= \theta_{12},
~~~ \bar\theta_{13} = \theta_{13},
~~~ \bar\theta_{14} = \theta_{14},
\end{equation}
\begin{eqnarray} \label{mutau1}
\sin{\bar\theta_{24}} &=&  \sin\theta_{34} \cos{\theta_{24}} \\ 
\sin {\bar\theta_{23}}
&=&\frac{\cos{\theta_{23}}\cos{\theta_{34}}-\sin{\theta_{23}}\sin{\theta_{34}}\sin{\theta_{24}}}{\sqrt{1-\cos{\theta_{24}^2}\sin{\theta_{34}^2}}} \\
\sin {\bar\theta_{34}} \label{mutau2}
&=&\frac{\sin{\theta_{24}}}{\sqrt{1-\cos{\theta_{24}^2}\sin{\theta_{34}^2}}}.
\end{eqnarray}
Due to these complex relations the  behaviour of $M_{e\mu}$  is different from that of 
$M_{e\tau}$ unlike in three active neutrino case
where the plots of these two elements were same except for $\theta_{23}$ 
which differed in octant for both the textures.\\
It is found that in the limit of small $\theta_{24}$ the two active sterile mixing angles $\bar\theta_{24} \approx \theta_{34}$ from Eq. (\ref{mutau1}). The same can be seen from Eq. (\ref{mutau2}) which gives
$\bar\theta_{34} \approx \theta_{24}$ for smaller values of the mixing angle $\theta_{34}$.
Thus, for small $\theta_{24}$ and $\theta_{34}$,
the behaviour shown by $\theta_{24}$ in $M_{e\mu}$ ($M_{\mu\mu}$) is same 
as shown by $\theta_{34}$ in $M_{e\tau}$ ($M_{\tau\tau}$).

In Fig. (\ref{figmetau}) we have have plotted the correlation plots for
 $|M_{e \tau}|=0$. The upper panels are for NH and middle panels are for IH.
As discussed above, from the plots we see that the properties shown by $s_{24}^2$ in $M_{e \mu}$ are similar for $s_{34}^2$ in this case. Here we can see that
for NH, the allowed values of $s_{34}^2$ are very small. Unlike $M_{e \mu}$, here $s_{14}^2$ is allowed in its complete $3\sigma$ range. As there is no lower limit on $\theta_{34}$,
it can be extremely small (of the order of $10^{-4}$) even when $s_{14}^2$ is large to give allowed texture. 
For IH, the result is also similar to that of $M_{e \mu}$, where we can see that for $\alpha = 0^\circ$, the higher values of $s_{14}^2 s_{34}^2$ are not preferred and for 
$\alpha = +90^\circ$, the very low values of $s_{14}^2 s_{34}^2$ are not allowed. We also notice that the values of $s_{14}^2 s_{34}^2$ are restricted within 
0.002 which is of the same order in magnitude as obtained for $s_{14}^2 s_{24}^2$ in $M_{e\mu}$.
The bounds for the effective mass $M_{ee}$ are obtained as $< 0.09$ eV for NH and $< 0.14$  eV for IH.

In the lower panels of Fig. (\ref{figmetau}), we give the correlation plots for $M_{e \tau}$ 
for quasidegenerate mass spectrum. 
As mentioned above, the behaviour of this texture is related to $M_{e\mu}$ by the $\mu - \tau$ permutation symmetry. 
From the lower left pane of Fig. (\ref{figmetau}), we see that the nature of correlation for ($m_4 \sin^2\theta_{34}$) in $|M_{e \tau}| = 0$, is exactly same as that of ($m_4\sin^2 \theta_{24}$)
in $|M_{e\mu}| = 0$ (lower left panel of Fig. (\ref{figmemuih})).

\begin{figure*}
\begin{center}
\includegraphics[width=0.47\textwidth]{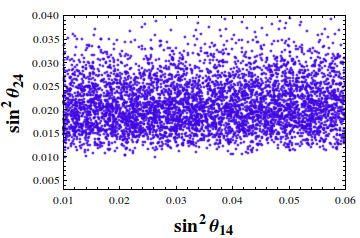}
\includegraphics[width=0.47\textwidth]{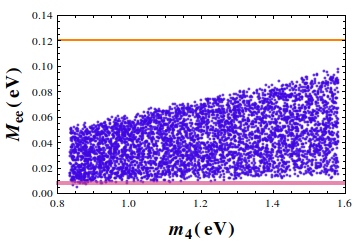}\\
\includegraphics[width=0.47\textwidth]{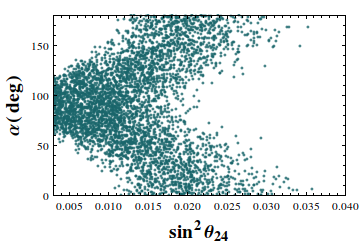}
\includegraphics[width=0.47\textwidth]{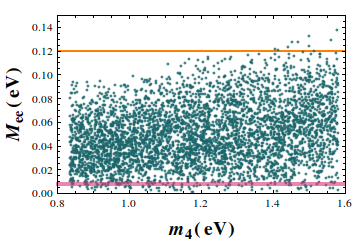}\\
\includegraphics[width=0.47\textwidth]{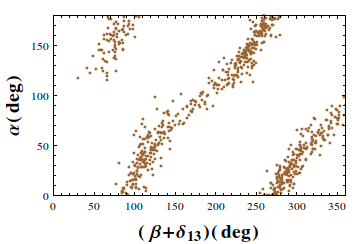}
\includegraphics[width=0.47\textwidth]{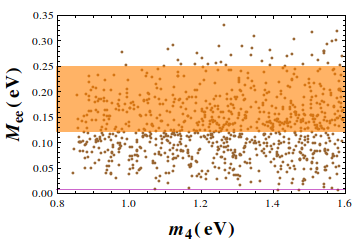}\\
\end{center}
\begin{center}
\caption{Correlation plots for $|M_{\mu \mu}|$ = 0 for NH (upper panel), IH (middle panel) and QD (lower panel) mass spectrum. Bounds on $M_{ee}$ are described in the caption of Fig.(\ref{figmemuih}).}
\label{figmmumu}
\end{center}
\end{figure*}

\subsection {The mass matrix elements $M_{\mu \mu}$ and $M_{\tau \tau}$}
%
The (2,2) diagonal matrix element in the 3+1 scenario is given as,
\bea \label{mmumu_text}
M_{\mu \mu} &=& e^{2 i(\delta_{14} - \delta_{24} + \gamma)} c_{14}^2 m_4 s_{24}^2 \\ \nonumber
&+& e^{2 i (\delta_{13} + \beta)} m_3(c_{13} c_{24} s_{23} \\ \nonumber
&-& e^{i(\delta_{14} - \delta_{13} -\delta_{24})} s_{13} s_{14} s_{24})^2 
+ m_1 \{-c_{23} c_{24} s_{12}  \\ \nonumber
&+& c_{12}(-e^{i \delta_{13}} c_{24} s_{13} s_{23} - e^{i(\delta_{14} - \delta_{24})} c_{13} s_{14} s_{24})\}^2 \\ \nonumber
&+& e^{2 i \alpha} m_2 \{c_{12} c_{23} c_{24} \\ \nonumber
&+& s_{12}(-e^{i \delta_{13}} c_{24} s_{13} s_{23} - e^{i(\delta_{14} - \delta_{24})} c_{13} s_{14} s_{24})\}^2 .
\eea
For NH, considering lowest vanishing mass ($m_1 = 0$), using the approximations as defined in Eqs. (\ref{xnh}, \ref{a1}) and putting the Majorana phases equal to zero as well as the Dirac phases equal to $180^\circ$, the zero texture condition yields:  
\bea \label{mmmphase}
&& s_{23}^2 + c_{12}^2 c_{23}^2 \sqrt{R_\nu} + c_{12} s_{12} \sin 2 \theta_{23}\sqrt{R_\nu} \lambda a_{13} \\ \nonumber
&+&\lambda^2(s_{12}^2s_{23}^2\sqrt{R_\nu}a_{13}^2-c_{23}\sin2\theta_{12}\sqrt{R_\nu}a_{14}a_{24} \\ \nonumber
&+&\sqrt{R_{\nu1}}a_{24}^2)=0.
\eea
From the expression we can understand that when $\theta_{24}$
is small, the leading order terms become larger as compared to the $R_{\nu1}$ term. 
On the other hand when $\theta_{24}$ is very large, the $R_{\nu1}$ becomes larger than the
leading order terms. Thus it will not be possible to have $|M_{\mu \mu}| = 0$ in NH for either very small or very large values of $\theta_{24}$. This is evident from the upper left panel of Fig. (\ref{figmmumu}). Here we can see that though $s_{14}^2$ is allowed in its $3\sigma$ range but the smaller as well as higher values of $\theta_{24}$ are disallowed. In this case the effective mass $M_{ee}$ is always less than 0.10 eV (upper right panel of  Fig. \ref{figmmumu}). 

For IH and $m_3 = 0$, using the approximations as defined in Eqs. (\ref{xih}, \ref{a1}) the expression for $M_{\mu \mu}$ becomes:
\bea
M_{\mu \mu} &\approx& c_{23}^2(s_{12}^2 + c_{12}^2 e^{2 i \alpha}) \\ \nonumber
&+& \frac{1}{2}
 \lambda \sin2\theta_{12} \sin2\theta_{23} e^{i \delta_{13}} ( 1 - e^{2 i \alpha}) a_{13} \\ \nonumber
 &+& \lambda^2[\sin2\theta_{12} c_{23} e^{i(\delta_{14} - \delta_{24})}(1 - e^{2 i \alpha}) a_{14} a_{24} \\ \nonumber
 &+& s_{23}^2 e^{2 i \delta_{13}}(c_{12}^2 + e^{ 2 i \alpha} s_{12}^2) a_{13}^2 \\ \nonumber
 &+& e^{2i(\gamma + \delta_{14} - \delta_{24})} \sqrt{R_{\nu1}} a_{24}^2 ].
\eea 
Further putting the phase $\alpha = 0^\circ$ and $\theta_{24} = 0^\circ$, 
the above equation simplifies to
\bea
M_{\mu \mu} \approx c_{23}^2 + s_{23}^2 s_{13}^2 e^{2 i \delta_{13}}
\eea
Thus $M_{\mu \mu}$ can not vanish for IH when the phase $\alpha$ is zero and $\theta_{24}$ is small. But for the higher values of $\theta_{24}$ the $R_{\nu1}$ term will start contributing and it becomes possible to have cancellation even if $\alpha = 0^\circ$. On the other hand when $\alpha = +90^\circ$, the contributions of the $\lambda$ and $\lambda^2$ terms come into play and thus one can have
cancellations for the smaller values of $\theta_{24}$. This can be seen from the left middle panel of Fig. (\ref{figmmumu}) where we can see that at $\alpha = 0^\circ$, the smaller values
of $s_{24}^2$ are disallowed and at $\alpha = +90^\circ$ the values  less than 0.022 of $s_{24}^2$ are allowed. In this case the upper bound of the effective mass is 0.14 eV as seen from the right middle panel of
Fig. (\ref{figmmumu}).
In the lower panels of Fig. (\ref{figmmumu}), we have given the correlation plots of $|M_{\mu \mu}| = 0$ for QD mass spectrum. To understand these correlations, we have expanded Eq. (\ref{mmumu_text}) using the approximations as given in Eqs. (\ref{a1}) and  (\ref{qd}) to obtain the following:
\bea \label{QD_pmumu}
 && |(c_{12}^2 c_{23}^2 e^{2 i \alpha} + c_{23}^2 s_{12}^2 + e^{2 i (\beta + \delta_{13})} s_{23}^2)m_0\\ \nonumber
 && -2(a_{13}c_{12} c_{23}e^{i \delta_{13}}(e^{2 i \alpha}-1)m_{0}s_{12}s_{23})\lambda + \mathcal{O}(\lambda^{2})|=0.
\eea
\begin{figure*}
\begin{center}
\includegraphics[width=0.47\textwidth]{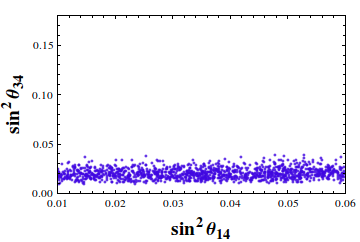}
\includegraphics[width=0.47\textwidth]{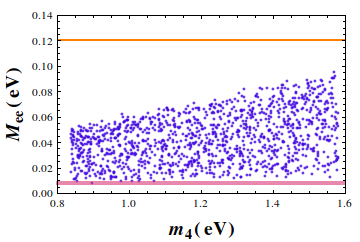}\\
\includegraphics[width=0.47\textwidth]{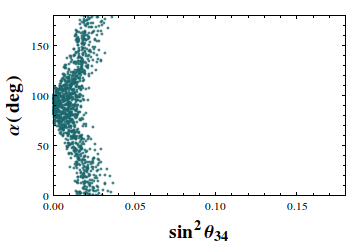}
\includegraphics[width=0.47\textwidth]{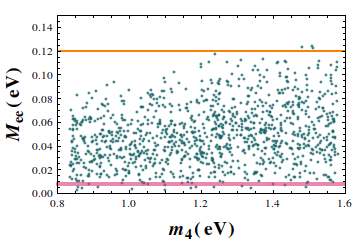}\\
\includegraphics[width=0.47\textwidth]{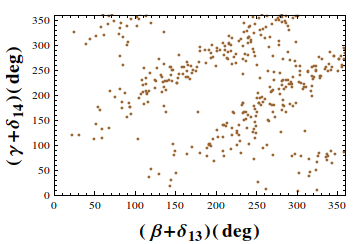}
\includegraphics[width=0.47\textwidth]{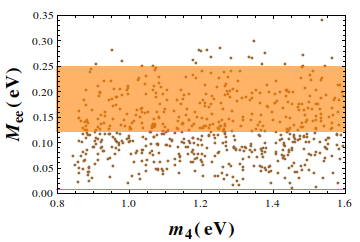}\\
\end{center}
\begin{center} 
\caption{Correlation plots for $|M_{\tau \tau}|$ = 0 for NH (upper panel), IH (middle panel) and QD (lower panel) mass spectrum. Bounds on $M_{ee}$ are described in the caption of Fig.(\ref{figmemuih}).}
\label{fig5ih}
\end{center}
\end{figure*} 
In this case we note that, unlike the textures $|M_{e \mu}|=0$ and $|M_{e \tau}|=0$, the factor
$ (e^{2 i\alpha}-1)$ appears with the subleading $\lambda$ term and the leading order term contains the  $(\beta + \delta_{13})$ factor. For the values $\alpha$ = 0$^\circ$ and $(\beta + \delta_{13})$ = 90$^\circ$, the survived terms of Eq. (\ref{QD_pmumu}) are $(m_0 \cos2\theta_{23})$ and the subleading $\lambda^2$ term. 
As both the terms have the magnitude of the order of $\mathcal{O}(10^{-2})$, there
can be cancellation of these terms leading to the possibility of vanishing $M_{\mu \mu}$.
However, for $\alpha$ = 90$^\circ$ and $(\beta + \delta_{13})$ = 90$^\circ$, the magnitude of the leading order term ($m_0 [-c_{23}^2 \cos2\theta_{12} s_{23}^2$]) is large ($\mathcal{O}(10^{-1})$)
than the subleading terms having coefficient $\lambda$ and $\lambda^2$. Thus one can not have cancellation at these values of $\alpha$ and $(\beta + \delta_{13})$. These conclusions
are clearly reflected in the lower left panel of Fig. (\ref{figmmumu}), where we have plotted $\alpha$ as a function of $(\beta + \delta_{13})$. 

The mass matrix element $M_{\tau \tau}$ in the 3+1 mixing scenario is given as,
\bea \label{mtautau_qd}
  M_{\tau \tau} &=& m_4 e^{2i(\delta_{14} + \gamma)} c_{14}^2 c_{24}^2 s_{34}^2 \\ \nonumber 
  &+&m_3 e^{2i(\delta_{13} + \beta)}  \{e^{i(\delta_{14} - \delta_{13})} c_{24} s_{13} s_{14} s_{34} \\ \nonumber
  &+&c_{13}(c_{23} c_{34} - e^{i \delta_{24}} s_{23} s_{24} s_{34})\}^2 \\ \nonumber
  &+& m_1[-s_{12}(-c_{34} s_{23} - e^{i \delta_{24}} c_{23} s_{24} s_{34}) \\ \nonumber
  &+& c_{12}\{-e^{i\delta_{14}} c_{13} c_{24} s_{14} s_{34} - e^{i \delta_{13}} s_{13}(c_{23} c_{34} \\ \nonumber
  &-& e^{i \delta_{24}} s_{23} s_{24} s_{34})\}]^2 \\ \nonumber
  &+& e^{2 i \alpha} m_2 [c_{12}(-c_{34} s_{23} - e^{i \delta_{24}} c_{23} s_{24} s_{34}) \\ \nonumber
  &+& s_{12}\{-e^{i \delta_{14}} c_{13} c_{24} s_{14} s_{34} - e^{i \delta_{13}} s_{13}(c_{23} c_{34} \\ \nonumber
  &-& e^{i \delta_{24}} s_{23} s_{24} s_{34})\}]^2
 \eea
As $M_{\tau \tau}$ is related to $ M_{\mu \mu}$ by $\mu-\tau$ symmetry, the behaviour of $\theta_{34}$ in $M_{\tau \tau}$ is similar to $\theta_{24}$ in $M_{\mu \mu}$.
This can be seen from the correlation plots for $|M_{\tau \tau}| = 0$ in Fig. (\ref{fig5ih}). 

For NH, very low and high values of $s_{34}^2$ are  disallowed (upper left panel of Fig. (\ref{fig5ih})) and 
for IH the correlation between the phase $\alpha$ and $s_{34}^2$ (middle left panel of Fig. (\ref{fig5ih}))
is exactly similar with the $\alpha$ and $s_{24}^2$ correlation in $M_{\mu \mu}$ and in both the cases $s_{24}^2$ and $s_{34}^2$ are restricted to below 0.04.
The upper bounds of the effective mass in this case is 0.10 eV (0.14 eV) in NH (IH) as shown in upper (middle) right panel of Fig. (\ref{fig5ih}).

For the QD case, using the approximation as given in Eqs. (\ref{qd}) and (\ref{a1}), Eq. (\ref{mtautau_qd}) can be simplified as,
\bea \label{tautauqd}
&&  | m_{0} ( c_{23}^{2} c_{34}^{2} e^{2 i (\beta + \delta_{13})} + 
   (c_{12}^{2} e^{2 i \alpha}+s_{12}^{2}) s_{23}^{2} c_{34}^{2} ) \\ \nonumber
&& + e^{2 i( \gamma + \delta_{14})} m_{4} s_{34}^{2})
 + \mathcal{O}(\lambda) +  \mathcal{O}(\lambda^{2})| = 0
\eea
Here, unlike $M_{\mu \mu}$, the phase factors  $(\beta + \delta_{13})$, $\alpha$\footnote{In this texture, we find that $\alpha$ is allowed in its full range.} and 
$(\gamma + \delta_{14})$ appear with the leading order $m_0$ term.
From the lower left panel of Fig. (\ref{fig5ih}), we see that in the ($(\beta + \delta_{13})$, $(\gamma + \delta_{14})$) plane ($ 0^\circ $, $ 0^\circ $) point is not allowed. 
To understand this point, we consider all phases to be zero in Eq. (\ref{tautauqd}) and this gives
\bea \label{tautauqd_ph0}
m_0 + (m_4 - m_0)s^{2}_{34} + (a^{2}_{14} + a^{2}_{24})(m_0 - m_4)s^{2}_{34} \lambda^{2}
\eea
Here we can see that the above equation is free from the $\lambda$ term. 
Thus cancellation between leading order 
term ($\mathcal{O} (10^{-1})$) and $ \lambda^{2}$ term ($\mathcal{O} (10^{-2})$) is not possible for any values of the remaining oscillation parameters. 
Whereas, for other allowed values of phases, for example: $ (\beta + \delta_{13}) = 250^\circ $ and $ ( \gamma + \delta_{14}) = 150^\circ $, 
the figure shows that one can have $|M_{\tau \tau}|=0$ for quasidegenerate mass spectrum.
%
\subsection {The mass matrix elements $M_{\mu \tau}$}
\begin{figure*}
\begin{center}
\includegraphics[width=0.47\textwidth]{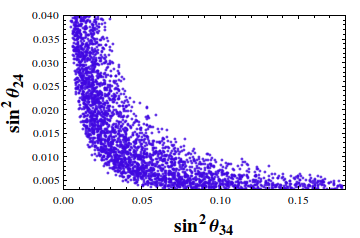}
\includegraphics[width=0.47\textwidth]{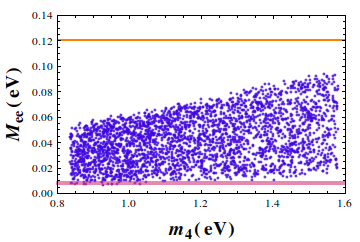}\\
\includegraphics[width=0.47\textwidth]{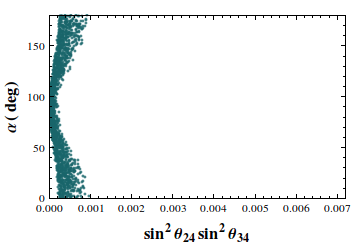}
\includegraphics[width=0.47\textwidth]{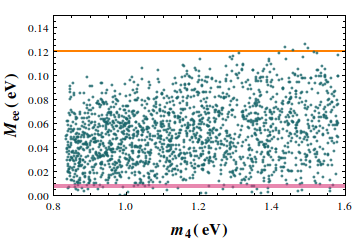}\\
\includegraphics[width=0.47\textwidth]{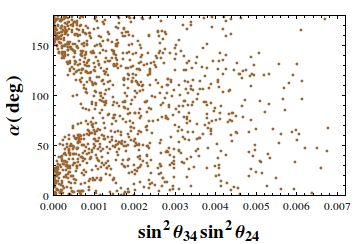}
\includegraphics[width=0.47\textwidth]{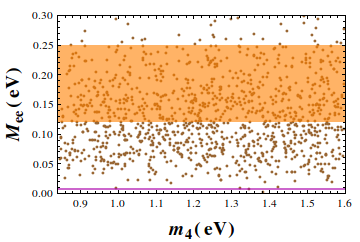}
\end{center}
\begin{center}
\caption{Correlation plots for $|M_{\mu \tau}|$ = 0 for NH (upper panel), IH (middle panel) and QD (lower panel) mass spectrum. Bounds on $M_{ee}$ are described in the caption of Fig.(\ref{figmemuih}).}
\label{fig8ih}
\end{center}
\end{figure*}
The full expression of $M_{\mu \tau}$ element in the 3+1 scenario is given by
\bea
 M_{\mu \tau} &=& e^{i(2 \delta_{14} - \delta_{24} + 2 \gamma)} c_{14}^2 c_{24} m_4 s_{24} s_{34} \\ \nonumber
 &+& e^{2 i (\delta_{13} + \beta)} m_3 ( c_{13} c_{24} s_{23}
 - e^{i(\delta_{14} - \delta_{24} - \delta_{13})} \\ \nonumber
 &&s_{13} s_{14} s_{24}) \{-e^{i(\delta_{14} - \delta_{13})} c_{24} s_{13} s_{14} s_{34} \\ \nonumber
 &+& c_{13} (c_{23} c_{34} - e^{i \delta_{24}} s_{23} s_{24} s_{34})\} + m_1\{- c_{23} c_{24} s_{12} \\ \nonumber 
 &+& c_{12}(-e^{i \delta_{13}} c_{24} s_{13} s_{23}- e^{i(\delta_{14} - \delta_{24})} c_{13} s_{14} s_{24})\} \\ \nonumber
 && [-s_{12} (-c_{34} s_{23} - e^{i \delta_{24}} c_{23} s_{24} s_{34}) + c_{12} \{ -e^{i \delta_{14}} c_{13} c_{24} \\ \nonumber
 && s_{14} s_{34} - e^{i \delta_{13}} s_{13}(c_{23} c_{34} - e^{i\delta_{24}} s_{23} s_{24} s_{34})\}] \\ \nonumber
 &+& e^{2 i \alpha} m_2\{c_{12} c_{23} c_{24} \\ \nonumber
 &+& s_{12}(-e^{i \delta_{13}} c_{24} s_{13} s_{23} - e^{i(\delta_{14} - \delta_{24})} c_{13} s_{14} s_{24})\} \\ \nonumber
 && [c_{12}(-c_{34} s_{23} - e^{i \delta_{24}} c_{23} s_{24} s_{34}) \\ \nonumber
 &+& s_{12}\{-e^{i \delta_{14}} c_{13} c_{24} s_{14} s_{34}  \\ \nonumber
 &-& e^{i \delta_{13}} s_{13}(c_{23} c_{34} - e^{i \delta_{24}} s_{23} s_{24} s_{34})\}].
\eea

For NH, when $m_1$ is zero, using the approximations as defined in  Eqs. (\ref{a1}, \ref{xnh}) and putting the Majorana phases equals to zero and the Dirac phases equals to
$180^\circ$, the above equation simplifies to
\bea
&&| c_{23} c_{34}s_{23}(1-c_{12}^2\sqrt{R_\nu}) \\ \nonumber
&+&\lambda\{(c_{12}c_{34}s_{12}\sqrt{R_\nu}a_{13})\cos2\theta_{23} \\ \nonumber
 &+& a_{24}s_{34}(s_{23}^2+c_{12}^2c_{23}^2\sqrt{R_\nu}) \\ \nonumber
 &+&s_{34}(\sqrt{R_{\nu 1}}a_{24}+c_{12}c_{23}s_{12}\sqrt{R_\nu}a_{14})\} \\ \nonumber
 &+& \lambda^2\{s_{12}a_{13}s_{23}s_{34}\sqrt{R_\nu}(s_{12}a_{14}+2c_{12}c_{23}a_{24}) \\ \nonumber
 &+& a_{14}s_{23}(c_{12}c_{34}s_{12}^2s_{23}\sqrt{R_\nu}a_{13}^2)\} |= 0.
\eea
From the above equation we can see that the $R_{\nu 1}$ appears with both
 $\theta_{14}$ and $\theta_{24}$. Thus cancellation of sterile term with
  the active neutrino terms will not be possible
when both these parameters are very large. 
This is evident from the upper left panel of Fig. (\ref{fig8ih}) from where we notice that when $s_{34}^2$ is small (high), $s_{24}^2$ is high (small).
But note that when both the angles are very small, then the active part becomes larger than the sterile term and cancellation is again not possible.
Here the bound of the effective mass $M_{ee}$ is $< 0.10$ eV (top right panel of Fig. (\ref{fig8ih})). 
\begin{figure*}
\begin{center}
\includegraphics[width=0.47\textwidth]{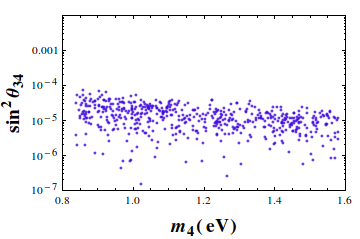}
\includegraphics[width=0.47\textwidth]{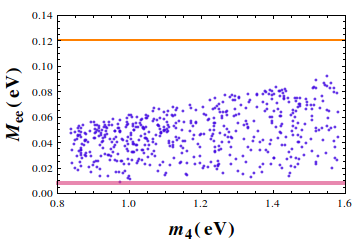}\\
\includegraphics[width=0.47\textwidth]{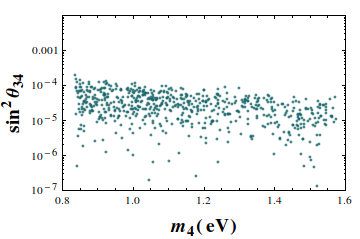}
\includegraphics[width=0.47\textwidth]{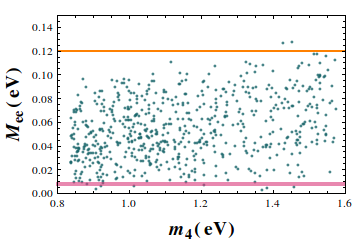}\\
\includegraphics[width=0.47\textwidth]{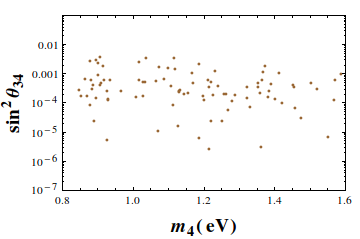}
\includegraphics[width=0.47\textwidth]{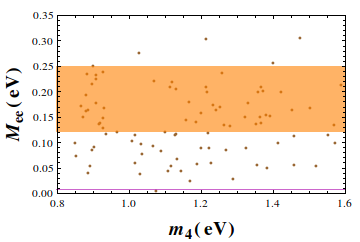}\\
\end{center}
\begin{center}
\caption{Correlation plots for $|M_{\tau s}|$ = 0 for NH (upper panel), IH (middle panel) and QD (lower panel) mass spectrum. Bounds on $M_{ee}$ are described in the caption of Fig.(\ref{figmemuih}).}
\label{fig10ih}
\end{center}
\end{figure*}
%
For IH when $m_3 = 0$, using the approximations of Eqs. (\ref{a1}, \ref{xnh})  the expression for $M_{\mu\tau}$ becomes,
\bea
 M_{\mu \tau} &\approx&\{ -c_{23} c_{34} s_{23}(c_{12}^2 e^{2 i \alpha} + s_{12}^2) \\ \nonumber
 &+& \lambda [c_{12} s_{12} (1 - e^{2 i \alpha})(c_{34} \cos2\theta_{23} e^{i \delta_{13}} a_{13}  \\ \nonumber
 &+& c_{23} s_{34} e^{i \delta_{14}} a_{14}) \\ \nonumber
 &+& s_{34}\{e^{i(2 \gamma+ 2 \delta_{14} - \delta_{24})} \sqrt{R_{\nu1}}  \\ \nonumber
 &-& c_{23}^2 e^{i \delta_{24}}(s_{12}^2 + c_{12}^2 e^{2 i \alpha})\}a_{24}] \\ \nonumber
 &+& \lambda^2 [c_{23} c_{34} s_{23} e^{ 2 i \delta_{13}}(c_{12}^2 + e^{2 i \alpha} s_{12}^2) a_{13}^2 \\ \nonumber
 &+& c_{12} s_{12} s_{23} (e^{2 i \alpha} - 1)(c_{34} e^{i(\delta_{14} - \delta_{24})} a_{14}  \\ \nonumber
 &+& 2 s_{34} c_{23} e^{i(\delta_{13} + \delta_{24})} a_{13})a_{24}]\}.
\eea
Taking $\alpha = 0^\circ$ and either $\theta_{34} = 0^\circ$ or $\theta_{24} = 0^\circ$ the above equation simplifies to
\bea
M_{\mu \tau} = -c_{23} c_{34} s_{23} (1 + s_{13}^2 e^{2 i \delta_{13}})
\label{mt}
\eea
Thus the texture $|M_{\mu \tau}|=0$ will be disallowed in IH for very small values of $\theta_{24}$ and $\theta_{34}$ when $\alpha = 0^\circ$ but will be allowed when the $R_{\nu1}$
term becomes of the same order as the term in Eq. (\ref{mt}). As when $\alpha$ approaches to $\pi/2$, the other terms start to contribute and thus one can have cancellations 
for small values of the above mentioned sterile mixing angles. This can be seen from the middle left panel of Fig. (\ref{fig8ih}). Here we notice that for $\alpha=0, \pi$
the allowed range of $s_{24}^2 s_{34}^2$ is (0.0002 - 0.001) and for $\alpha=\pi/2$, we obtain  $s_{24}^2 s_{34}^2 < 0.0004$. In this case the effective mass has an upper bound of 0.013 eV.

The correlation plots for $|M_{\mu \tau}|$ = 0 for QD neutrino mass spectrum are given in the
lower panels of Fig (\ref{fig8ih}). The expression for this texture is quite complicated and to understand the correlations between parameters we take the Majorana phases $\alpha$ = 90$^\circ$ and ($\beta + \delta_{13}$) = 0$^\circ$ and vanishing Dirac phases $\delta_{14}$ and $\delta_{24}$. These approximations gives the expression for zero texture at $M_{\mu\tau}$ as,
\bea \nonumber
&& |\sin 2\theta_{23}c_{12}^2 c_{34} m_0 + [-a_{24}s_{34}(m_0 (s_{23}^2 - c_{23}^2 \cos 2\theta_{12})\\ \nonumber
 && -m_4)+m_0 \sin 2\theta_{12}(a_{13}e^{i\delta_{13}} \cos 2\theta_{23} +a_{14}c_{23}s_{34})]\lambda\\
 && + \mathcal{O}(\lambda^{2})|=0.
\eea
The leading term ($\mathcal{O}$ (10$^{-2}$)) in this case is always greater than the subleading terms 
$\mathcal{O} (10^{-3})$ for very small values of $s_{24} s_{34}$. Therefore there is no allowed solution when $\alpha = 0^\circ$ and $\theta_{14}$ and $\theta_{24}$ are very small. On the other hand when we increase the value of $s_{24} s_{34}$ the subleading term with coefficient $\lambda$ becomes sufficiently large and gets cancelled with the leading order term to give allowed points. These conclusions are quite clearly depicted
in the lower left panel of Fig. (\ref{fig8ih}).

\subsection {The mass matrix element $M_{\tau s}$}

The full expression for the element $M_{\tau s}$ is given by,
\bea   \label{mtaus}
  M_{\tau s} &=& c_{14}^2 c_{24}^2 c_{34} e^{ 2 i(\delta_{14} + \gamma)} m_4 s_{34} \\ \nonumber
  &+&  (m_1, m_2, m_3) { \rm terms}.
 \eea

We note that the $m_4$ term appears with $s_{34}$. Thus it is clear that if $\theta_{34}$ is very high then the sterile term will be very large and it is impossible to have cancellations with the active terms.

The texture $|M_{\tau s}|=0$ in both NH and IH will be only possible for very small values of $\theta_{34}$.
This has been shown in the upper left and  middle left panels of Fig. (\ref{fig10ih}). In both the plots we note that as $m_4$ increases, the value of $s_{34}^2$ decreases.
For both the hierarchies, the highest possible value of $s_{34}^2$ for which cancellation can occur is of the order of $10^{-4}$. Here the effective mass $M_{ee}$
is restricted within 0.10 eV (0.14 eV) in NH (IH) (upper (middle) right panels of Fig. (\ref{fig10ih})).

$|M_{\tau s}| = 0$ for QD is marginally allowed in the region $10^{-6} < s_{34}^2 < 0.01$ as reflected in 
the lower left panel of Fig (\ref {fig10ih}). In this case, we note that the allowed values of $\theta_{34}$ are slightly higher as compared to the NH and IH case.
This is because in the QD regime, the contributions from active terms are greater as compared to the completely hierarchical regime. Thus to have cancellation, the $m_4$ terms needs to be large,
and that is why higher values of $\theta_{34}$ are preferred in this case.

\subsection {The mass matrix elements $M_{es}$, $M_{\mu s}$ and $M_{ss}$}

The zero textures at matrix elements $M_{es}$, $M_{\mu s}$ and $M_{ss}$ are disallowed by the current data if we assume that the lowest mass vanishes.
This can be understood by looking at the $m_4$ terms of these elements.
\bea \label{mmus}
M_{es} &=& c_{14} c_{24} c_{34} e^{ 2 i(\delta_{14} + \gamma)} m_4 s_{14} \\ \nonumber
  &+&  (m_1, m_2, m_3) { \rm terms} ,\\  \nonumber
M_{\mu s} &=& c_{14}^2 c_{24} c_{34} e^{ 2 i(\delta_{14} + \gamma)} m_4 s_{24} \\ \nonumber
  &+&  (m_1, m_2, m_3) { \rm terms} ,\\ \nonumber
M_{ss} &=& c_{14}^2 c_{24}^2 c_{34}^2 e^{ 2 i(\delta_{14} + \gamma)} m_4 \\ \nonumber
  &+&  (m_1, m_2, m_3) { \rm terms}. \nonumber
\eea
In the elements $M_{es}$ and $M_{\mu s}$ we see that the $m_4$ term appears with $s_{14}$ and $s_{24}$ respectively. Thus these terms are in general large and 
can be cancelled with the active terms only for the higher values of the lowest masses ($m_1$ for NH and $m_3$ for IH). Because when the lowest mass vanishes,
the active terms become smaller than the $m_4$ term and thus cancellation is not possible. For $M_{ss}$, we see that the $m_4$ term is not suppressed
by the sterile mixing angles
and thus it is of the order of $\sim 1$ eV. For this reason, it is never possible to have $|M_{ss}|=0$ for NH, IH and QD mass spectrum.

For the QD scenario, the condition for texture zero for $M_{es}$ is given by, 
\bea \label{QD_mes}
&& |c_{12} s_{23} s_{34} s_{12} m_0 (e^{2 i \alpha}-1) + (a_{14}c_{34}e^{i\delta_{14}}(m_0(c_{12}^2+\\ \nonumber
&& s_{12}^2 e^{2i\alpha})-m_4e^{2i\gamma}) -m_0 c_{23}(a_{24}c_{12}c_{34}e^{i\delta_{24}}(e^{2i\alpha}-1)s_{12} \\ \nonumber
&& +a_{13}e^{i\delta_{13}}(-(c_{12}^2 +s_{12}^2 e^{2 i \alpha})+e^{2 i \beta})s_{34}))) \lambda
+ \mathcal{O}(\lambda^{2})| =0 \\ \nonumber 
\eea
By looking at the above expression we can conclude that even when active neutrinos are quasidegenerate, it is not possible to have $|M_{es}|=0$.
This is because the term $m_4 \sin \theta_{14}$ which appears with a coefficient $\lambda$, has far greater magnitude ($\sim$ 10$^{-1}$) than the other two
terms.

\begin{figure*}
\begin{center}
\includegraphics[width=0.47\textwidth]{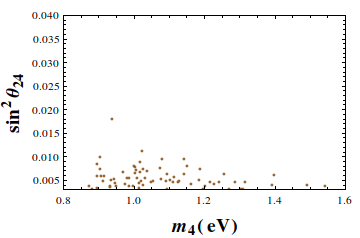}
\includegraphics[width=0.47\textwidth]{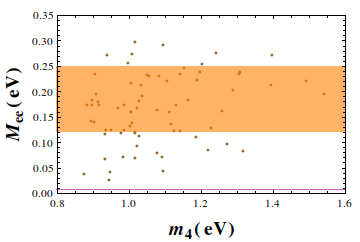}\\
\end{center}
\begin{center}
\caption{Correlation plots for $|M_{\mu s}|$ = 0 for QD mass spectrum. Band represents the
allowed range of $M_{ee}$ are described in the caption of Fig.(\ref{figmemuih}).}
\label{fig11ih}
\end{center}
\end{figure*}

Regarding $|M_{\mu s}|$ we find that, this element can marginally vanish in the QD mass spectrum. Using the approximations for QD and Eq. (\ref{a1}) the
zero texture condition of $M_{\mu s}$ translates to: 
\bea \label{QD_mmus}
&& |-c_{23} s_{23} s_{34} m_0 (-c^{2}_{12}e^{2 i \alpha}+ e^{2 i (\beta + \delta_{13})}- s^{2}_{12}) + \\ \nonumber
&& e^{- i \delta_{24}}(-a_{24}c_{34}(c^{2}_{12} c^{2}_{23}e^{2 i (\alpha + \delta_{24})}m_{0}  
- e^{2 i (\gamma + \delta_{14})}m_{4} \\ \nonumber
&& + c^{2}_{23}s^{2}_{12} e^{2 i \delta_{24}}m_{0} + s^{2}_{23} e^{2 i (\beta +\delta_{13}+ \delta_{24})} s^{2}_{23} m_{0}    ) \\ \nonumber
&&  c_{12}s_{12} e^{i \delta_{24}}(e^{2 i \alpha} -1 )m_{0} (a_{14}c_{23}c_{34}e^{i \delta_{14}} \\ \nonumber
&& - a_{13}e^{i \delta_{13}}(c_{23}^2-s_{23}^2)s_{34}) )\lambda + \mathcal{O}(\lambda^{2})| =0
\eea
We have given the correlation plots for this texture in Fig.  (\ref{fig11ih}). 
Taking all the Majorana and the Dirac phases to be zero, we find that 
for the smaller values of $\theta_{24}$, all the terms in the above expression
are of the same order ($\mathcal{O}$ (10$^{-1}$)) and thus there are viable solutions. This can be
seen from the left panel of Fig (\ref{fig11ih}).  
%
%

\section{Conclusions}
\label{sec5}

 We present the phenomenological analysis of one-zero texture of 
$4 \times 4$ low energy neutrino mass matrix ($M_\nu$) in presence of an 
extra sterile neutrino in addition to the three active ones (3+1 scenario).
To analyse the behaviour of active sterile mixing parameters we consider two different scenarios:
(i) completely hierarchical mass spectrum of active neutrinos with vanishing lowest mass (i.e., NH and IH)
(ii) completely quasidegenerate active neutrino masses where all the active neutrinos have almost equal mass (i.e., QD).  This symmetric matrix $M_\nu$ has 10 independent entries and thus in general have 10 possible
structures with one-zero texture. We extensively study the viability and correlation between different parameters of all these textures for both the cases using the neutrino oscillation data 
for active neutrinos \cite{Gonzalez-Garcia:2015qrr,Forero:2014bxa,Capozzi:2013csa} and the SBL data for 
active sterile mixing \cite{Kopp:2013vaa,schwetz}. For completely hierarchical mass spectrum only 7 texture zero structures are consistent with the current data and 8 are consistent for quasidegenerate mass spectrum.
In Table (\ref{allow_tab}), we have summarize the allowed and disallowed one-zero textures in neutrino mass matrix for both the mass spectrum under consideration. To understand our numerical results, we expand the matrix elements in terms of the small parameters by keeping the terms upto second order.
We observe that the predictions of one-zero texture mass matrices change significantly when the cases corresponding to vanishing lowest active neutrino mass and quasidegenerate mass spectrum are compared. 	
%
\begin{table}
\centering
\begin{tabular}{|| c | c | c | c ||}
\hline
M$_{\alpha \beta = 0}$ &  NH($ m_{1} =0 $) & IH($ m_{3} =0 $) & QD \\
\hline
$ M_{ee} $  &  $ \times $ &  $\surd$ &  $\surd$   \\
$ M_{e\mu} $  &   $\surd$ &  $\surd$ &  $\surd$   \\
$ M_{e\tau} $  &  $\surd$ &  $\surd$ &  $\surd$   \\
$ M_{\mu \tau} $  &  $\surd$ &  $\surd$ &  $\surd$   \\
$ M_{\mu \mu} $  &  $\surd$ &  $\surd$ &  $\surd$   \\
$ M_{\tau \tau} $  &  $\surd$ &  $\surd$ &  $\surd$   \\
$ M_{es} $  &  $ \times $ &  $ \times $ &  $ \times $   \\
$ M_{\mu s} $  &  $ \times $ &  $ \times $ &  $\surd$   \\
$ M_{\tau s} $  &  $\surd$ &  $\surd$ &  $\surd$   \\
$ M_{s s} $  &  $ \times $ &  $ \times $ &  $ \times $ \\
\hline
\end{tabular}
\begin{center}
\caption{Possible zero-textures that are allowed (disallowed) and these are marked with $\surd$ ($ \times $) for the three mass patterns when the parameters are varied in their allowed $3 \sigma$ range.}
\label{allow_tab}
\end{center}
\end{table}
%
The texture $|M_{ee}|=0$ is allowed only for IH and QD.
The Majorana phase $\gamma$ is highly constrained around +$ 90^\circ $ for
both IH and QD, on the other hand, the other phase $\alpha$ is constrained only for QD masses.
In 3-neutrino scenario, the element $ M_{e \mu} $ and $ M_{e \tau} $  are related by simple $ \mu - \tau $ permutation symmetry but in 3+1 scenario, the relations are very complex as compared to the three generation case. Here the behaviour of $\theta_{24}$ in $M_{e\mu}$ is quite similar with the behaviour of $\theta_{34}$ in $M_{e \tau}$. We find that the texture $ |M_{e \mu}|=0$ mainly constrains the active sterile mixing angles 
$ \theta_{14} $, $ \theta_{24} $ for NH and IH while for QD mass spectrum, the Majorana phase $\alpha$ is largely constrained for smaller values of $\theta_{24}$. For vanishing $ M_{e \tau}$ very small values of active sterile mixing angle $ \theta_{34}$ is allowed for NH and IH. For QD spectrum Majorana phase $\alpha$ is constrained for small values of $\theta_{34}$.
The textures $ |M_{\mu \mu}| = 0$ and $ |M_{\tau \tau}| = 0 $ are again related by $ \mu - \tau $ symmetry and thus the characteristics of $ \theta_{24}$ in $M_{\mu \mu}$ and $\theta_{34}$ in $M_{\tau \tau}$ are same.
For $|M_{\mu \mu}| = 0$ and NH we find that very low and high values of 
$ s^{2}_{24} $ are disallowed whereas for $ |M_{\tau \tau}| = 0$ 
very low and  high values of $ s^{2}_{34} $ are disallowed. For IH we observe that for lower values of $ s^{2}_{24} $ ($ s^{2}_{34} $) most of the  values of the 
Majorana phase $ \alpha $ is disallowed in case of texture $ M_{\mu \mu} $ ($M_{\tau \tau} $). For
QD, however complete range of this angle is allowed for both these textures.
For $| M_{\mu \tau}| = 0$ to be phenomenologically viable, the mixing angle $ \theta_{24} $ and $ \theta_{34} $ can not be very large  or small simultaneously for NH.
For IH the behaviour of $ \alpha$ vs $s^{2}_{24} s^{2}_{34} $ is  same as that of $ \alpha$ vs $s^{2}_{24}$ ($ \alpha$ vs $s^{2}_{34}$) in $ M_{\mu \mu} $($ M_{\tau \tau} $). 
For QD, $\alpha = +90^\circ$ is not allowed for very small values of $\theta_{24}$ and $\theta_{34}$.
For $M_{\tau s}$ to vanish, the parameter $ s^{2}_{34} $ should be as low as $ 10^{-4} $ irrespective of the hierarchy (NH or IH). However, for QD slightly higher values of $\theta_{34}$ are preferred.
The remaining three textures $M_{es}$, $M_{\mu s}$ and $M_{ss}$ are phenomenologically disallowed when 
we consider completely hierarchical mass spectrum with a vanishing neutrino mass.
The sterile contribution in the first two cases $\propto$ $m_4 s_{14}$ and $m_4 s_{24}$ respectively. These terms are in general large and there is no cancellation possible for these terms when the 
lowest mass is vanishing.
However, for QD $|M_{\mu s}| = 0$ is marginally allowed for very small values of $\theta_{24}$. 
The texture, $M_{ss}$ is of the order of $\mathcal{O}$ (1) eV and hence
it is not possible to get contributions of this order from the active sector for NH (IH) and QD mass spectrum.

The results discussed in the above paragraph are obtained by varying the neutrino oscillation parameters in their allowed $3 \sigma$ range. However, we know that there are several current and
future experiments which are expected to put a more stringent constrain on the active-sterile mixing parameters \cite{An:2016luf,MINOS:2016viw,Adamson:2016jku}. 
To check what happens to allowed textures if the mixing angles are constrained furthermore,
we re-evaluate the viability of the textures by varying the oscillation parameters in their allowed $1 \sigma$ and $2 \sigma$ allowed range as given in Table \ref{parameters}.
We find that the texture $M_{\mu s}$ is disallowed for QD when the parameters
 are varied within their $1 \sigma$ range and the same is marginally allowed
  when the parameters are varied in their $2 \sigma$ range. 
Apart from this, the other conclusions remain unaltered.

For all allowed textures in completely hierarchical and quasidegenerate scenarios we obtain the bounds on the effective mass $M_{ee}$ that will be constrained by the forthcoming
neutrinoless double beta decay experiments. 
For NH, the values of $M_{ee}$ are well below the present bound and for IH it is within the present upper bound. In all the allowed textures we see that the upper bound on $M_{ee}$
for NH is slightly smaller as compared to IH. For QD mass spectrum the some of the allowed region of the parameter space can be discarded from the bounds available on $M_{ee}$. The similar behaviours of $M_{ee}$ for NH, IH and QD for all the allowed textures can be qualitatively understood from the following arguments. 
For NH, the $m_2$ and $m_3$ terms are proportional to the $\Delta m_{21}^2$ and 
$\Delta m_{31}^2$ respectively. On the other hand for IH and QD, all the active masses $m_i$ ($i=1, 2$ for IH and $i=1, 2, 3$ for QD) are proportional to  $\Delta m_{31}^2$ and  $ 0.1 - 0.3$ eV$^2$ respectively.
Thus we see that for NH, due to smaller values of the $m_2$ term, the
 prediction for $M_{ee}$ is lower. Whereas for IH and QD, $M_{ee}$ gets
  contribution from the large values of $m_i$ and thus the prediction 
of $M_{ee}$ is high for IH and even higher for QD as compared to NH.

If future experiments confirm the existence of a light sterile neutrino then the next challenge will be
to build models for such scenarios. The texture analysis performed  in this paper can be a useful
guide for constructing models for light sterile neutrinos. 

\section{Acknowledgements} 
The authors would like to thank Srubabati Goswami for her careful reading of the manuscript and helpful suggestions. They would also like to thank Ujjal Dey for useful discussions.
MG would like to thank Abhay Swain for help in Jaxodraw. The work of SG is supported by the Australian Research 
Council through the ARC Center of Excellence in Particle Physics (CoEPP Adelaide) at the Terascale (CE110001004).

\bibliography{one_zero}

\begin{thebibliography}{51}
\expandafter\ifx\csname natexlab\endcsname\relax\def\natexlab#1{#1}\fi
\expandafter\ifx\csname bibnamefont\endcsname\relax
  \def\bibnamefont#1{#1}\fi
\expandafter\ifx\csname bibfnamefont\endcsname\relax
  \def\bibfnamefont#1{#1}\fi
\expandafter\ifx\csname citenamefont\endcsname\relax
  \def\citenamefont#1{#1}\fi
\expandafter\ifx\csname url\endcsname\relax
  \def\url#1{\texttt{#1}}\fi
\expandafter\ifx\csname urlprefix\endcsname\relax\def\urlprefix{URL }\fi
\providecommand{\bibinfo}[2]{#2}
\providecommand{\eprint}[2][]{\url{#2}}

\bibitem[{\citenamefont{Gonzalez-Garcia
  et~al.}(2015)\citenamefont{Gonzalez-Garcia, Maltoni, and
  Schwetz}}]{Gonzalez-Garcia:2015qrr}
\bibinfo{author}{\bibfnamefont{M.~C.} \bibnamefont{Gonzalez-Garcia}},
  \bibinfo{author}{\bibfnamefont{M.}~\bibnamefont{Maltoni}}, \bibnamefont{and}
  \bibinfo{author}{\bibfnamefont{T.}~\bibnamefont{Schwetz}}
  (\bibinfo{year}{2015}), \eprint{1512.06856}.

\bibitem[{\citenamefont{Forero et~al.}(2014)\citenamefont{Forero, Tortola, and
  Valle}}]{Forero:2014bxa}
\bibinfo{author}{\bibfnamefont{D.~V.} \bibnamefont{Forero}},
  \bibinfo{author}{\bibfnamefont{M.}~\bibnamefont{Tortola}}, \bibnamefont{and}
  \bibinfo{author}{\bibfnamefont{J.~W.~F.} \bibnamefont{Valle}},
  \bibinfo{journal}{Phys. Rev.} \textbf{\bibinfo{volume}{D90}},
  \bibinfo{pages}{093006} (\bibinfo{year}{2014}), \eprint{1405.7540}.

\bibitem[{\citenamefont{Capozzi et~al.}(2014)\citenamefont{Capozzi, Fogli,
  Lisi, Marrone, Montanino, and Palazzo}}]{Capozzi:2013csa}
\bibinfo{author}{\bibfnamefont{F.}~\bibnamefont{Capozzi}},
  \bibinfo{author}{\bibfnamefont{G.~L.} \bibnamefont{Fogli}},
  \bibinfo{author}{\bibfnamefont{E.}~\bibnamefont{Lisi}},
  \bibinfo{author}{\bibfnamefont{A.}~\bibnamefont{Marrone}},
  \bibinfo{author}{\bibfnamefont{D.}~\bibnamefont{Montanino}},
  \bibnamefont{and} \bibinfo{author}{\bibfnamefont{A.}~\bibnamefont{Palazzo}},
  \bibinfo{journal}{Phys. Rev.} \textbf{\bibinfo{volume}{D89}},
  \bibinfo{pages}{093018} (\bibinfo{year}{2014}), \eprint{1312.2878}.

\bibitem[{\citenamefont{Athanassopoulos et~al.}(1996)}]{Athanassopoulos:1996jb}
\bibinfo{author}{\bibfnamefont{C.}~\bibnamefont{Athanassopoulos}}
  \bibnamefont{et~al.} (\bibinfo{collaboration}{LSND}),
  \bibinfo{journal}{Phys.Rev.Lett.} \textbf{\bibinfo{volume}{77}},
  \bibinfo{pages}{3082} (\bibinfo{year}{1996}), \eprint{nucl-ex/9605003}.

\bibitem[{\citenamefont{Athanassopoulos et~al.}(1998)}]{Athanassopoulos:1997pv}
\bibinfo{author}{\bibfnamefont{C.}~\bibnamefont{Athanassopoulos}}
  \bibnamefont{et~al.} (\bibinfo{collaboration}{LSND}),
  \bibinfo{journal}{Phys.Rev.Lett.} \textbf{\bibinfo{volume}{81}},
  \bibinfo{pages}{1774} (\bibinfo{year}{1998}), \eprint{nucl-ex/9709006}.

\bibitem[{\citenamefont{Aguilar-Arevalo et~al.}(2001)}]{Aguilar:2001ty}
\bibinfo{author}{\bibfnamefont{A.}~\bibnamefont{Aguilar-Arevalo}}
  \bibnamefont{et~al.} (\bibinfo{collaboration}{LSND}),
  \bibinfo{journal}{Phys.Rev.} \textbf{\bibinfo{volume}{D64}},
  \bibinfo{pages}{112007} (\bibinfo{year}{2001}), \eprint{hep-ex/0104049}.

\bibitem[{\citenamefont{Aguilar-Arevalo
  et~al.}(2013)}]{Aguilar-Arevalo:2013pmq}
\bibinfo{author}{\bibfnamefont{A.~A.} \bibnamefont{Aguilar-Arevalo}}
  \bibnamefont{et~al.} (\bibinfo{collaboration}{MiniBooNE}),
  \bibinfo{journal}{Phys. Rev. Lett.} \textbf{\bibinfo{volume}{110}},
  \bibinfo{pages}{161801} (\bibinfo{year}{2013}), \eprint{1207.4809}.

\bibitem[{\citenamefont{Mention et~al.}(2011)\citenamefont{Mention, Fechner,
  Lasserre, Mueller, Lhuillier et~al.}}]{Mention:2011rk}
\bibinfo{author}{\bibfnamefont{G.}~\bibnamefont{Mention}},
  \bibinfo{author}{\bibfnamefont{M.}~\bibnamefont{Fechner}},
  \bibinfo{author}{\bibfnamefont{T.}~\bibnamefont{Lasserre}},
  \bibinfo{author}{\bibfnamefont{T.}~\bibnamefont{Mueller}},
  \bibinfo{author}{\bibfnamefont{D.}~\bibnamefont{Lhuillier}},
  \bibnamefont{et~al.}, \bibinfo{journal}{Phys.Rev.}
  \textbf{\bibinfo{volume}{D83}}, \bibinfo{pages}{073006}
  (\bibinfo{year}{2011}), \eprint{1101.2755}.

\bibitem[{\citenamefont{Giunti and
  Laveder}(2011{\natexlab{a}})}]{Giunti:2010zu}
\bibinfo{author}{\bibfnamefont{C.}~\bibnamefont{Giunti}} \bibnamefont{and}
  \bibinfo{author}{\bibfnamefont{M.}~\bibnamefont{Laveder}},
  \bibinfo{journal}{Phys.Rev.} \textbf{\bibinfo{volume}{C83}},
  \bibinfo{pages}{065504} (\bibinfo{year}{2011}{\natexlab{a}}),
  \eprint{1006.3244}.

\bibitem[{\citenamefont{Ade et~al.}(2015)}]{Ade:2015xua}
\bibinfo{author}{\bibfnamefont{P.~A.~R.} \bibnamefont{Ade}}
  \bibnamefont{et~al.} (\bibinfo{collaboration}{Planck})
  (\bibinfo{year}{2015}), \eprint{1502.01589}.

\bibitem[{\citenamefont{Frampton et~al.}(2002)\citenamefont{Frampton, Glashow,
  and Marfatia}}]{Frampton:2002yf}
\bibinfo{author}{\bibfnamefont{P.~H.} \bibnamefont{Frampton}},
  \bibinfo{author}{\bibfnamefont{S.~L.} \bibnamefont{Glashow}},
  \bibnamefont{and} \bibinfo{author}{\bibfnamefont{D.}~\bibnamefont{Marfatia}},
  \bibinfo{journal}{Phys.Lett.} \textbf{\bibinfo{volume}{B536}},
  \bibinfo{pages}{79} (\bibinfo{year}{2002}), \eprint{hep-ph/0201008}.

\bibitem[{\citenamefont{Dev et~al.}(2007{\natexlab{a}})\citenamefont{Dev,
  Kumar, Verma, and Gupta}}]{Dev:2006qe}
\bibinfo{author}{\bibfnamefont{S.}~\bibnamefont{Dev}},
  \bibinfo{author}{\bibfnamefont{S.}~\bibnamefont{Kumar}},
  \bibinfo{author}{\bibfnamefont{S.}~\bibnamefont{Verma}}, \bibnamefont{and}
  \bibinfo{author}{\bibfnamefont{S.}~\bibnamefont{Gupta}},
  \bibinfo{journal}{Phys.Rev.} \textbf{\bibinfo{volume}{D76}},
  \bibinfo{pages}{013002} (\bibinfo{year}{2007}{\natexlab{a}}),
  \eprint{hep-ph/0612102}.

\bibitem[{\citenamefont{Xing}(2002{\natexlab{a}})}]{Xing:2002ta}
\bibinfo{author}{\bibfnamefont{Z.-z.} \bibnamefont{Xing}},
  \bibinfo{journal}{Phys.Lett.} \textbf{\bibinfo{volume}{B530}},
  \bibinfo{pages}{159} (\bibinfo{year}{2002}{\natexlab{a}}),
  \eprint{hep-ph/0201151}.

\bibitem[{\citenamefont{Xing}(2002{\natexlab{b}})}]{Xing:2002ap}
\bibinfo{author}{\bibfnamefont{Z.-z.} \bibnamefont{Xing}},
  \bibinfo{journal}{Phys.Lett.} \textbf{\bibinfo{volume}{B539}},
  \bibinfo{pages}{85} (\bibinfo{year}{2002}{\natexlab{b}}),
  \eprint{hep-ph/0205032}.

\bibitem[{\citenamefont{Desai et~al.}(2003)\citenamefont{Desai, Roy, and
  Vaucher}}]{Desai:2002sz}
\bibinfo{author}{\bibfnamefont{B.~R.} \bibnamefont{Desai}},
  \bibinfo{author}{\bibfnamefont{D.}~\bibnamefont{Roy}}, \bibnamefont{and}
  \bibinfo{author}{\bibfnamefont{A.~R.} \bibnamefont{Vaucher}},
  \bibinfo{journal}{Mod.Phys.Lett.} \textbf{\bibinfo{volume}{A18}},
  \bibinfo{pages}{1355} (\bibinfo{year}{2003}), \eprint{hep-ph/0209035}.

\bibitem[{\citenamefont{Dev et~al.}(2007{\natexlab{b}})\citenamefont{Dev,
  Kumar, Verma, and Gupta}}]{Dev:2007fs}
\bibinfo{author}{\bibfnamefont{S.}~\bibnamefont{Dev}},
  \bibinfo{author}{\bibfnamefont{S.}~\bibnamefont{Kumar}},
  \bibinfo{author}{\bibfnamefont{S.}~\bibnamefont{Verma}}, \bibnamefont{and}
  \bibinfo{author}{\bibfnamefont{S.}~\bibnamefont{Gupta}},
  \bibinfo{journal}{Phys.Lett.} \textbf{\bibinfo{volume}{B656}},
  \bibinfo{pages}{79} (\bibinfo{year}{2007}{\natexlab{b}}), \eprint{0708.3321}.

\bibitem[{\citenamefont{Dev et~al.}(2007{\natexlab{c}})\citenamefont{Dev,
  Kumar, Verma, and Gupta}}]{Dev:2006xu}
\bibinfo{author}{\bibfnamefont{S.}~\bibnamefont{Dev}},
  \bibinfo{author}{\bibfnamefont{S.}~\bibnamefont{Kumar}},
  \bibinfo{author}{\bibfnamefont{S.}~\bibnamefont{Verma}}, \bibnamefont{and}
  \bibinfo{author}{\bibfnamefont{S.}~\bibnamefont{Gupta}},
  \bibinfo{journal}{Nucl.Phys.} \textbf{\bibinfo{volume}{B784}},
  \bibinfo{pages}{103} (\bibinfo{year}{2007}{\natexlab{c}}),
  \eprint{hep-ph/0611313}.

\bibitem[{\citenamefont{Kumar}(2011)}]{Kumar:2011vf}
\bibinfo{author}{\bibfnamefont{S.}~\bibnamefont{Kumar}},
  \bibinfo{journal}{Phys.Rev.} \textbf{\bibinfo{volume}{D84}},
  \bibinfo{pages}{077301} (\bibinfo{year}{2011}), \eprint{1108.2137}.

\bibitem[{\citenamefont{Fritzsch et~al.}(2011)\citenamefont{Fritzsch, Xing, and
  Zhou}}]{Fritzsch:2011qv}
\bibinfo{author}{\bibfnamefont{H.}~\bibnamefont{Fritzsch}},
  \bibinfo{author}{\bibfnamefont{Z.-z.} \bibnamefont{Xing}}, \bibnamefont{and}
  \bibinfo{author}{\bibfnamefont{S.}~\bibnamefont{Zhou}},
  \bibinfo{journal}{JHEP} \textbf{\bibinfo{volume}{1109}}, \bibinfo{pages}{083}
  (\bibinfo{year}{2011}), \eprint{1108.4534}.

\bibitem[{\citenamefont{Meloni and Blankenburg}(2013)}]{Meloni:2012sx}
\bibinfo{author}{\bibfnamefont{D.}~\bibnamefont{Meloni}} \bibnamefont{and}
  \bibinfo{author}{\bibfnamefont{G.}~\bibnamefont{Blankenburg}},
  \bibinfo{journal}{Nucl.Phys.} \textbf{\bibinfo{volume}{B867}},
  \bibinfo{pages}{749} (\bibinfo{year}{2013}), \eprint{1204.2706}.

\bibitem[{\citenamefont{Ludl et~al.}(2012)\citenamefont{Ludl, Morisi, and
  Peinado}}]{Ludl:2011vv}
\bibinfo{author}{\bibfnamefont{P.}~\bibnamefont{Ludl}},
  \bibinfo{author}{\bibfnamefont{S.}~\bibnamefont{Morisi}}, \bibnamefont{and}
  \bibinfo{author}{\bibfnamefont{E.}~\bibnamefont{Peinado}},
  \bibinfo{journal}{Nucl.Phys.} \textbf{\bibinfo{volume}{B857}},
  \bibinfo{pages}{411} (\bibinfo{year}{2012}), \eprint{1109.3393}.

\bibitem[{\citenamefont{Grimus and Ludl}(2013)}]{Grimus:2012zm}
\bibinfo{author}{\bibfnamefont{W.}~\bibnamefont{Grimus}} \bibnamefont{and}
  \bibinfo{author}{\bibfnamefont{P.}~\bibnamefont{Ludl}},
  \bibinfo{journal}{J.Phys.} \textbf{\bibinfo{volume}{G40}},
  \bibinfo{pages}{055003} (\bibinfo{year}{2013}), \eprint{1208.4515}.

\bibitem[{\citenamefont{Grimus et~al.}(2004)\citenamefont{Grimus, Joshipura,
  Lavoura, and Tanimoto}}]{Grimus:2004hf}
\bibinfo{author}{\bibfnamefont{W.}~\bibnamefont{Grimus}},
  \bibinfo{author}{\bibfnamefont{A.~S.} \bibnamefont{Joshipura}},
  \bibinfo{author}{\bibfnamefont{L.}~\bibnamefont{Lavoura}}, \bibnamefont{and}
  \bibinfo{author}{\bibfnamefont{M.}~\bibnamefont{Tanimoto}},
  \bibinfo{journal}{Eur. Phys. J.} \textbf{\bibinfo{volume}{C36}},
  \bibinfo{pages}{227} (\bibinfo{year}{2004}), \eprint{hep-ph/0405016}.

\bibitem[{\citenamefont{Low}(2005)}]{Low:2005yc}
\bibinfo{author}{\bibfnamefont{C.~I.} \bibnamefont{Low}},
  \bibinfo{journal}{Phys. Rev.} \textbf{\bibinfo{volume}{D71}},
  \bibinfo{pages}{073007} (\bibinfo{year}{2005}), \eprint{hep-ph/0501251}.

\bibitem[{\citenamefont{Dev et~al.}(2011)\citenamefont{Dev, Gupta, and
  Gautam}}]{Dev:2011jc}
\bibinfo{author}{\bibfnamefont{S.}~\bibnamefont{Dev}},
  \bibinfo{author}{\bibfnamefont{S.}~\bibnamefont{Gupta}}, \bibnamefont{and}
  \bibinfo{author}{\bibfnamefont{R.~R.} \bibnamefont{Gautam}},
  \bibinfo{journal}{Phys. Lett.} \textbf{\bibinfo{volume}{B701}},
  \bibinfo{pages}{605} (\bibinfo{year}{2011}), \eprint{1106.3451}.

\bibitem[{\citenamefont{Ghosh et~al.}(2013{\natexlab{a}})\citenamefont{Ghosh,
  Goswami, and Gupta}}]{Ghosh:2012pw}
\bibinfo{author}{\bibfnamefont{M.}~\bibnamefont{Ghosh}},
  \bibinfo{author}{\bibfnamefont{S.}~\bibnamefont{Goswami}}, \bibnamefont{and}
  \bibinfo{author}{\bibfnamefont{S.}~\bibnamefont{Gupta}},
  \bibinfo{journal}{JHEP} \textbf{\bibinfo{volume}{04}}, \bibinfo{pages}{103}
  (\bibinfo{year}{2013}{\natexlab{a}}), \eprint{1211.0118}.

\bibitem[{\citenamefont{Ghosh et~al.}(2013{\natexlab{b}})\citenamefont{Ghosh,
  Goswami, Gupta, and Kim}}]{Ghosh:2013nya}
\bibinfo{author}{\bibfnamefont{M.}~\bibnamefont{Ghosh}},
  \bibinfo{author}{\bibfnamefont{S.}~\bibnamefont{Goswami}},
  \bibinfo{author}{\bibfnamefont{S.}~\bibnamefont{Gupta}}, \bibnamefont{and}
  \bibinfo{author}{\bibfnamefont{C.~S.} \bibnamefont{Kim}},
  \bibinfo{journal}{Phys. Rev.} \textbf{\bibinfo{volume}{D88}},
  \bibinfo{pages}{033009} (\bibinfo{year}{2013}{\natexlab{b}}),
  \eprint{1305.0180}.

\bibitem[{\citenamefont{Zhang}(2013)}]{Zhang:2013mb}
\bibinfo{author}{\bibfnamefont{Y.}~\bibnamefont{Zhang}},
  \bibinfo{journal}{Phys.Rev.} \textbf{\bibinfo{volume}{D87}},
  \bibinfo{pages}{053020} (\bibinfo{year}{2013}), \eprint{1301.7302}.

\bibitem[{\citenamefont{Lashin and Chamoun}(2012)}]{Lashin:2011dn}
\bibinfo{author}{\bibfnamefont{E.~I.} \bibnamefont{Lashin}} \bibnamefont{and}
  \bibinfo{author}{\bibfnamefont{N.}~\bibnamefont{Chamoun}},
  \bibinfo{journal}{Phys. Rev.} \textbf{\bibinfo{volume}{D85}},
  \bibinfo{pages}{113011} (\bibinfo{year}{2012}), \eprint{1108.4010}.

\bibitem[{\citenamefont{Barry et~al.}(2011)\citenamefont{Barry, Rodejohann, and
  Zhang}}]{Barry:2011wb}
\bibinfo{author}{\bibfnamefont{J.}~\bibnamefont{Barry}},
  \bibinfo{author}{\bibfnamefont{W.}~\bibnamefont{Rodejohann}},
  \bibnamefont{and} \bibinfo{author}{\bibfnamefont{H.}~\bibnamefont{Zhang}},
  \bibinfo{journal}{JHEP} \textbf{\bibinfo{volume}{1107}}, \bibinfo{pages}{091}
  (\bibinfo{year}{2011}), \eprint{1105.3911}.

\bibitem[{\citenamefont{Zhang}(2012)}]{Zhang:2011vh}
\bibinfo{author}{\bibfnamefont{H.}~\bibnamefont{Zhang}},
  \bibinfo{journal}{Phys. Lett.} \textbf{\bibinfo{volume}{B714}},
  \bibinfo{pages}{262} (\bibinfo{year}{2012}), \eprint{1110.6838}.

\bibitem[{\citenamefont{Bhupal~Dev and Pilaftsis}(2013)}]{Dev:2012bd}
\bibinfo{author}{\bibfnamefont{P.~S.} \bibnamefont{Bhupal~Dev}}
  \bibnamefont{and}
  \bibinfo{author}{\bibfnamefont{A.}~\bibnamefont{Pilaftsis}},
  \bibinfo{journal}{Phys. Rev.} \textbf{\bibinfo{volume}{D87}},
  \bibinfo{pages}{053007} (\bibinfo{year}{2013}), \eprint{1212.3808}.

\bibitem[{\citenamefont{Giunti and
  Laveder}(2011{\natexlab{b}})}]{Giunti:2011gz}
\bibinfo{author}{\bibfnamefont{C.}~\bibnamefont{Giunti}} \bibnamefont{and}
  \bibinfo{author}{\bibfnamefont{M.}~\bibnamefont{Laveder}},
  \bibinfo{journal}{Phys. Rev.} \textbf{\bibinfo{volume}{D84}},
  \bibinfo{pages}{073008} (\bibinfo{year}{2011}{\natexlab{b}}),
  \eprint{1107.1452}.

\bibitem[{\citenamefont{Schwetz}(2011)}]{schwetz}
\bibinfo{author}{\bibfnamefont{T.}~\bibnamefont{Schwetz}}
  (\bibinfo{year}{2011}), \bibinfo{note}{talk given at Proceedings of Sterile
  Neutrino Crossroads, 2011, Virginia Tech, USA}.

\bibitem[{\citenamefont{Goswami and Rodejohann}(2006)}]{Goswami:2005ng}
\bibinfo{author}{\bibfnamefont{S.}~\bibnamefont{Goswami}} \bibnamefont{and}
  \bibinfo{author}{\bibfnamefont{W.}~\bibnamefont{Rodejohann}},
  \bibinfo{journal}{Phys.Rev.} \textbf{\bibinfo{volume}{D73}},
  \bibinfo{pages}{113003} (\bibinfo{year}{2006}), \eprint{hep-ph/0512234}.

\bibitem[{\citenamefont{Kopp et~al.}(2013)\citenamefont{Kopp, Machado, Maltoni,
  and Schwetz}}]{Kopp:2013vaa}
\bibinfo{author}{\bibfnamefont{J.}~\bibnamefont{Kopp}},
  \bibinfo{author}{\bibfnamefont{P.~A.~N.} \bibnamefont{Machado}},
  \bibinfo{author}{\bibfnamefont{M.}~\bibnamefont{Maltoni}}, \bibnamefont{and}
  \bibinfo{author}{\bibfnamefont{T.}~\bibnamefont{Schwetz}},
  \bibinfo{journal}{JHEP} \textbf{\bibinfo{volume}{05}}, \bibinfo{pages}{050}
  (\bibinfo{year}{2013}), \eprint{1303.3011}.

\bibitem[{\citenamefont{Gorla}(2012)}]{Gorla:2012gd}
\bibinfo{author}{\bibfnamefont{P.}~\bibnamefont{Gorla}}
  (\bibinfo{collaboration}{CUORE}), \bibinfo{journal}{J.Phys.Conf.Ser.}
  \textbf{\bibinfo{volume}{375}}, \bibinfo{pages}{042013}
  (\bibinfo{year}{2012}).

\bibitem[{\citenamefont{Wilkerson et~al.}(2012)\citenamefont{Wilkerson, Aguayo,
  Avignone, Back, Barabash et~al.}}]{Wilkerson:2012ga}
\bibinfo{author}{\bibfnamefont{J.}~\bibnamefont{Wilkerson}},
  \bibinfo{author}{\bibfnamefont{E.}~\bibnamefont{Aguayo}},
  \bibinfo{author}{\bibfnamefont{F.}~\bibnamefont{Avignone}},
  \bibinfo{author}{\bibfnamefont{H.}~\bibnamefont{Back}},
  \bibinfo{author}{\bibfnamefont{A.}~\bibnamefont{Barabash}},
  \bibnamefont{et~al.}, \bibinfo{journal}{J.Phys.Conf.Ser.}
  \textbf{\bibinfo{volume}{375}}, \bibinfo{pages}{042010}
  (\bibinfo{year}{2012}).

\bibitem[{\citenamefont{Barabash}(2012)}]{Barabash:2012gc}
\bibinfo{author}{\bibfnamefont{A.}~\bibnamefont{Barabash}}
  (\bibinfo{collaboration}{SuperNEMO}), \bibinfo{journal}{J.Phys.Conf.Ser.}
  \textbf{\bibinfo{volume}{375}}, \bibinfo{pages}{042012}
  (\bibinfo{year}{2012}).

\bibitem[{\citenamefont{Gando et~al.}(2013)}]{Gando:2012zm}
\bibinfo{author}{\bibfnamefont{A.}~\bibnamefont{Gando}} \bibnamefont{et~al.}
  (\bibinfo{collaboration}{KamLAND-Zen}), \bibinfo{journal}{Phys.Rev.Lett.}
  \textbf{\bibinfo{volume}{110}}, \bibinfo{pages}{062502}
  (\bibinfo{year}{2013}), \eprint{1211.3863}.

\bibitem[{\citenamefont{Auger et~al.}(2012)}]{Auger:2012ar}
\bibinfo{author}{\bibfnamefont{M.}~\bibnamefont{Auger}} \bibnamefont{et~al.}
  (\bibinfo{collaboration}{EXO}), \bibinfo{journal}{Phys.Rev.Lett.}
  \textbf{\bibinfo{volume}{109}}, \bibinfo{pages}{032505}
  (\bibinfo{year}{2012}), \eprint{1205.5608}.

\bibitem[{\citenamefont{Dell'Oro et~al.}()\citenamefont{Dell'Oro, Marcocci, and
  Vissani}}]{DellOro:2014yca}
\bibinfo{author}{\bibfnamefont{S.}~\bibnamefont{Dell'Oro}},
  \bibinfo{author}{\bibfnamefont{S.}~\bibnamefont{Marcocci}}, \bibnamefont{and}
  \bibinfo{author}{\bibfnamefont{F.}~\bibnamefont{Vissani}},
  \bibinfo{journal}{Phys.Rev.}  (????).

\bibitem[{\citenamefont{Merle and Rodejohann}(2006)}]{Merle:2006du}
\bibinfo{author}{\bibfnamefont{A.}~\bibnamefont{Merle}} \bibnamefont{and}
  \bibinfo{author}{\bibfnamefont{W.}~\bibnamefont{Rodejohann}},
  \bibinfo{journal}{Phys. Rev.} \textbf{\bibinfo{volume}{D73}},
  \bibinfo{pages}{073012} (\bibinfo{year}{2006}), \eprint{hep-ph/0603111}.

\bibitem[{\citenamefont{Gautam et~al.}(2015)\citenamefont{Gautam, Singh, and
  Gupta}}]{Gautam:2015kya}
\bibinfo{author}{\bibfnamefont{R.~R.} \bibnamefont{Gautam}},
  \bibinfo{author}{\bibfnamefont{M.}~\bibnamefont{Singh}}, \bibnamefont{and}
  \bibinfo{author}{\bibfnamefont{M.}~\bibnamefont{Gupta}},
  \bibinfo{journal}{Phys. Rev.} \textbf{\bibinfo{volume}{D92}},
  \bibinfo{pages}{013006} (\bibinfo{year}{2015}), \eprint{1506.04868}.

\bibitem[{\citenamefont{Lam}(2005)}]{Lam:2005va}
\bibinfo{author}{\bibfnamefont{C.~S.} \bibnamefont{Lam}},
  \bibinfo{journal}{Phys. Rev.} \textbf{\bibinfo{volume}{D71}},
  \bibinfo{pages}{093001} (\bibinfo{year}{2005}), \eprint{hep-ph/0503159}.

\bibitem[{\citenamefont{Grimus and Lavoura}(2001)}]{Grimus:2001uc}
\bibinfo{author}{\bibfnamefont{W.}~\bibnamefont{Grimus}} \bibnamefont{and}
  \bibinfo{author}{\bibfnamefont{L.}~\bibnamefont{Lavoura}},
  \bibinfo{journal}{Acta Phys. Polon.} \textbf{\bibinfo{volume}{B32}},
  \bibinfo{pages}{3719} (\bibinfo{year}{2001}), \bibinfo{note}{[Acta
  Astron.32,3719(2001)]}, \eprint{hep-ph/0110041}.

\bibitem[{\citenamefont{Mohapatra and Rodejohann}(2005)}]{Mohapatra:2005yu}
\bibinfo{author}{\bibfnamefont{R.~N.} \bibnamefont{Mohapatra}}
  \bibnamefont{and}
  \bibinfo{author}{\bibfnamefont{W.}~\bibnamefont{Rodejohann}},
  \bibinfo{journal}{Phys. Rev.} \textbf{\bibinfo{volume}{D72}},
  \bibinfo{pages}{053001} (\bibinfo{year}{2005}), \eprint{hep-ph/0507312}.

\bibitem[{\citenamefont{Gupta et~al.}(2013)\citenamefont{Gupta, Joshipura, and
  Patel}}]{Gupta:2013it}
\bibinfo{author}{\bibfnamefont{S.}~\bibnamefont{Gupta}},
  \bibinfo{author}{\bibfnamefont{A.~S.} \bibnamefont{Joshipura}},
  \bibnamefont{and} \bibinfo{author}{\bibfnamefont{K.~M.} \bibnamefont{Patel}},
  \bibinfo{journal}{JHEP} \textbf{\bibinfo{volume}{09}}, \bibinfo{pages}{035}
  (\bibinfo{year}{2013}), \eprint{1301.7130}.

\bibitem[{\citenamefont{An et~al.}(2016)}]{An:2016luf}
\bibinfo{author}{\bibfnamefont{F.~P.} \bibnamefont{An}} \bibnamefont{et~al.}
  (\bibinfo{collaboration}{Daya Bay}) (\bibinfo{year}{2016}),
  \eprint{1607.01174}.

\bibitem[{\citenamefont{Adamson et~al.}(2016{\natexlab{a}})}]{MINOS:2016viw}
\bibinfo{author}{\bibfnamefont{P.}~\bibnamefont{Adamson}} \bibnamefont{et~al.}
  (\bibinfo{collaboration}{MINOS}) (\bibinfo{year}{2016}{\natexlab{a}}),
  \eprint{1607.01176}.

\bibitem[{\citenamefont{Adamson et~al.}(2016{\natexlab{b}})}]{Adamson:2016jku}
\bibinfo{author}{\bibfnamefont{P.}~\bibnamefont{Adamson}} \bibnamefont{et~al.}
  (\bibinfo{collaboration}{MINOS, Daya Bay}), \bibinfo{journal}{Submitted to:
  Phys. Rev. Lett.}  (\bibinfo{year}{2016}{\natexlab{b}}), \eprint{1607.01177}.

\end{thebibliography}

\end{document}